\documentclass[sigplan,10pt]{acmart}
\renewcommand\footnotetextcopyrightpermission[1]{}
\usepackage{mathptmx}

\usepackage{graphicx}
\usepackage[algoruled,vlined,linesnumbered]{algorithm2e}

\usepackage{amsmath,amsfonts,amssymb}
\usepackage{listings}
\usepackage{tabularx}
\usepackage{multicol}
\usepackage{float}

\usepackage{verbatim}
\usepackage[font=small, labelfont=bf, justification=justified]{caption}
\usepackage{subfigure}

\usepackage{alltt}
\usepackage{wrapfig}
\usepackage[mmddyyyy]{datetime}

\usepackage[T1]{fontenc}

\definecolor{codegreen}{rgb}{0,0.6,0}
\definecolor{codegray}{rgb}{0.5,0.5,0.5}
\definecolor{codepurple}{rgb}{0.58,0,0.82}
\definecolor{backcolour}{rgb}{0.95,0.95,0.92}
\definecolor{purple}{RGB}{128,0,128}
\definecolor{indigo}{RGB}{75,0,130}
\definecolor{royalblue}{RGB}{65,105,225}
\definecolor{navy}{RGB}{0,0,128}

\newcommand{\system}{InferLine\xspace}

\newcommand{\profiler}{\textbf{\small{Profiler}}\xspace}
\newcommand{\estimator}{\textbf{\small{Estimator}}\xspace}
\newcommand{\planner}{\textbf{\small{Planner}}\xspace}
\newcommand{\reactive}{\textbf{\small{Tuner}}\xspace}

\newif\ifcommenton
\commentontrue
\ifcommenton
\newcommand{\joey}[1]{\textcolor{blue}{\textit{(#1 - Joey)}}}
\newcommand{\dan}[1]{\textcolor{red}{\textit{(#1 - Dan)}}}
\newcommand{\alexey}[1]{\textcolor{indigo}{\textit{(#1 - AT)}}}
\newcommand{\ion}[1]{\textcolor{green}{\textit{(#1 - Ion)}}}
\newcommand{\eyal}[1]{\textcolor{teal}{\textit{(#1 - Eyal)}}}
\else
\newcommand{\alexey}[1]{}
\newcommand{\dan}[1]{}
\newcommand{\joey}[1]{}
\newcommand{\ion}[1]{}
\newcommand{\eyal}[1]{}
\fi

\newcommand{\eg}{{e.g.,}~}

\newcommand{\figref}[1]{Fig.~\ref{#1}}

\newcommand{\secref}[1]{\S\ref{#1}}

\newcommand{\algref}[1]{Algorithm~\ref{#1}}

\newcommand{\ceil}[1]{\left\lceil #1 \right\rceil}

\begin{document}
\frenchspacing

\title{\system: ML Prediction Pipeline Provisioning and Management for Tight Latency Objectives}
\author{Daniel Crankshaw, Gur-Eyal Sela, Simon Mo, Corey Zumar}
\author{Joseph E. Gonzalez, Ion Stoica, Alexey Tumanov}

\begin{abstract}
Serving ML prediction pipelines spanning multiple models and hardware accelerators is 
a key challenge in production machine learning.
Optimally configuring these pipelines to meet tight end-to-end latency goals is complicated by the interaction between model batch size, the choice of hardware accelerator, and variation in the query arrival process.

In this paper we introduce
\system, a system which provisions and manages the individual stages of prediction pipelines to meet end-to-end tail latency constraints while minimizing  cost. 
\system consists of a low-frequency combinatorial planner and a high-frequency auto-scaling tuner.
The low-frequency planner leverages stage-wise profiling, discrete event simulation, and constrained combinatorial search to automatically select hardware type, replication, and batching parameters for each stage in the pipeline. 
The high-frequency tuner uses network calculus to auto-scale each stage to meet tail latency goals in response to changes in the query arrival process.
We demonstrate that \system outperforms existing approaches by up to 7.6x in cost while achieving up to 34.5x lower latency SLO miss rate on realistic workloads and generalizes across state-of-the-art model serving frameworks.

\end{abstract}

\settopmatter{printfolios=true}
\maketitle


\section{Introduction}
\label{sec:intro}

Applications today increasingly rely on ML inference over multiple models linked together in a dataflow DAG.
Examples include a digital assistant service (e.g., Amazon Alexa), which combines audio pre-processing 
with downstream models for speech recognition, topic identification, question interpretation and response
and text-to-speech to answer a user's question.
The natural evolution of these applications leads to a growth in the complexity of the prediction pipelines.
At the same time, their latency-sensitive nature dictates tight tail latency constraints (e.g., 200-300ms).
As the pipelines grow and the models used become increasingly sophisticated, they present a unique set of systems challenges for provisioning and managing these pipelines.

\begin{figure}[tb]
	\centering
	\includegraphics[width=\columnwidth]{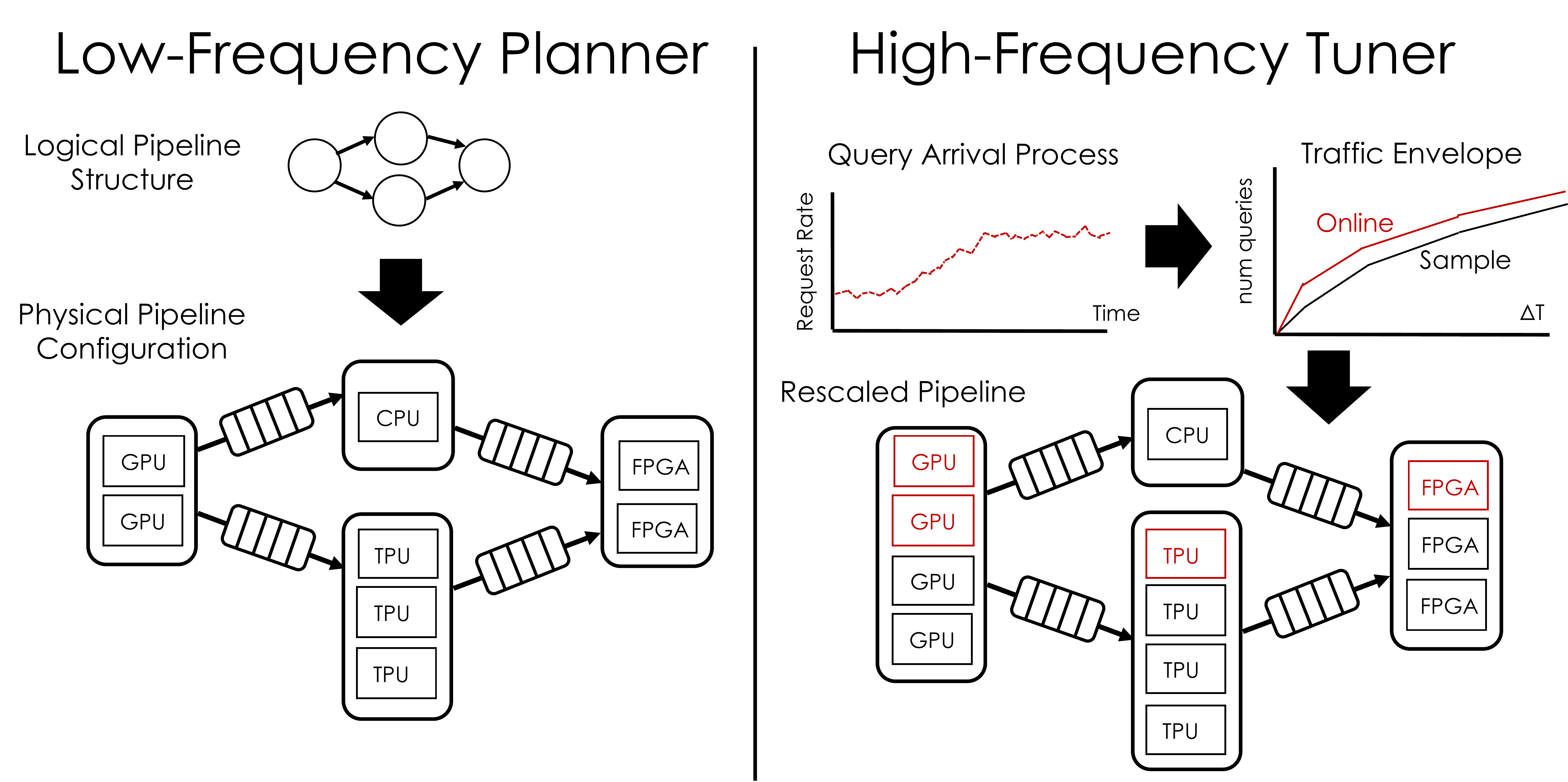}
	\vspace{-0.2in}
	\caption{
		\footnotesize{
		    \textbf{\system System Overview}
		}
	}
	\label{fig:sysarch:arch}
	\vspace{-0.2in}
\end{figure}

Each stage of the pipeline must be assigned the appropriate hardware accelerator (e.g., CPU, GPU, TPU) --- a task complicated
by increasing hardware heterogeneity.
Each model must be configured with the appropriate query batch size --- necessary for 
optimal utilization of the hardware.
And each pipeline stage can be replicated to meet the application throughput requirements.
Per-stage decisions with respect to the hardware type and batch size affect the latency contributed by each stage towards the end-to-end 
pipeline latency bound by the application-specified Service Level Objective (SLO).
This creates a combinatorial search space with three control dimensions per model (hardware type, batch size, number of replicas) and global constraints on aggregate latency.

A number of prediction serving systems exist today, including Clipper~\cite{clippernsdi17}, TensorFlow Serving~\cite{tfserving}, and NVIDIA TensorRT Inference Server~\cite{tensorrtserver} that optimize for single model serving. This pushes the complexity of coordinating cross-model interaction and, particularly, the questions of per-model configuration to meet application-level requirements, to the application developer.
To the best of our knowledge, no system exists today that automates the process of pipeline provisioning and configuration, subject to specified tail latency SLO in a cost-aware manner.
Thus, the goal of this paper is to address the problem of configuring and managing  multi-stage prediction pipelines subject to end-to-end tail latency constraints cost efficiently.

We propose \system{} --- a system for provisioning and management of ML inference pipelines. It composes with existing prediction serving frameworks, such as Clipper and TensorFlow Serving.
It is necessary for such a system to contain two principal components: a low-frequency \textit{planner} and a high-frequency \textit{tuner}. The low-frequency planner is responsible for navigating the combinatorial search space to produce per-model pipeline configuration relatively infrequently to minimize cost. It is intended to run periodically to correct for workload drift or fundamental changes in the steady-state, long-term query arrival process. It is also necessary for integrating new models added to the repository and to integrate new hardware accelerators. The high frequency component is intended to operate at time scales three orders of magnitude faster. It monitors instantaneous query arrival traffic and tunes the running pipeline to accomodate unexpected query spikes cost efficiently to maintain latency SLOs under bursty and stochastic workloads.


To enable efficient exploration of the combinatorial configuration space, 
\system profiles each stage in the pipeline individually and uses these profiles
and a discrete event simulator to accurately estimate end-to-end pipeline latency given the hardware configuration and batchsize parameters.
The low-frequency \emph{planner} uses a constrained greedy search algorithm to find the cost-minimizing pipeline configuration that meets the end-to-end tail latency constraint determined using the discrete event simulator on a sample planning trace.


The \system high-frequency \textit{tuner} leverages traffic envelopes built using  network calculus tools to capture the arrival process dynamics across multiple time scales and determine when and how to react to changes in the arrival process.
As a consequence, the tuner is able to maintain the latency SLO in the presence of transient spikes and sustained variation in the query arrival process.




In summary, the primary contribution of this paper is a system for provisioning and managing machine learning inference pipelines for latency-sensitive applications cost efficiently.
It consists of two principal components that operate at time scales orders of magnitude apart to configure the system for near-optimal performance. 
The planner builds on a high-fidelity model-based networked queueing simulator, while the tuner uses network calculus techniques to rapidly adjust pipeline configuration, absorbing unexpected query traffic variation cost efficiently.

We apply \system to provision and manage resources for multiple state-of-the-art prediction serving systems.  We show that \system significantly outperforms alternative pipeline configuration baselines by a factor of up to 7.6X on cost, while exceeding 99\% latency SLO attainment---the highest level of attainment achieved in relevant prediction serving literature.

\section{Background and Motivation}
\label{motivation}

\begin{figure*}[th]
	\begin{center}
		\subfigure[Image Processing]{\includegraphics[width=0.23\textwidth]{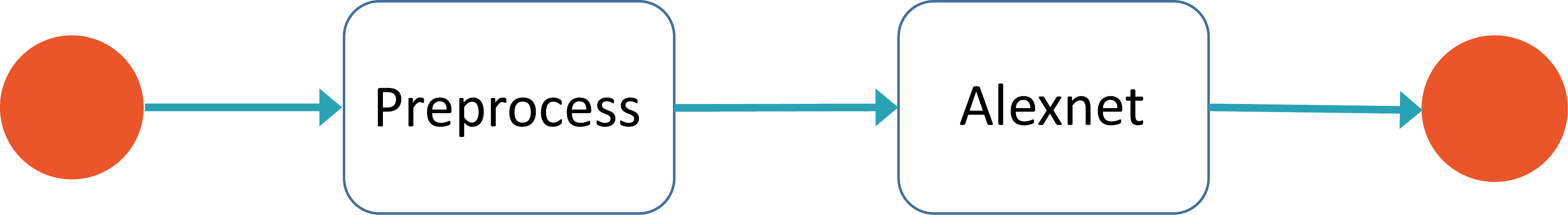}\label{fig:exp:heavycpu}
		}
		\subfigure[Video Monitoring]{\includegraphics[width=0.23\textwidth]{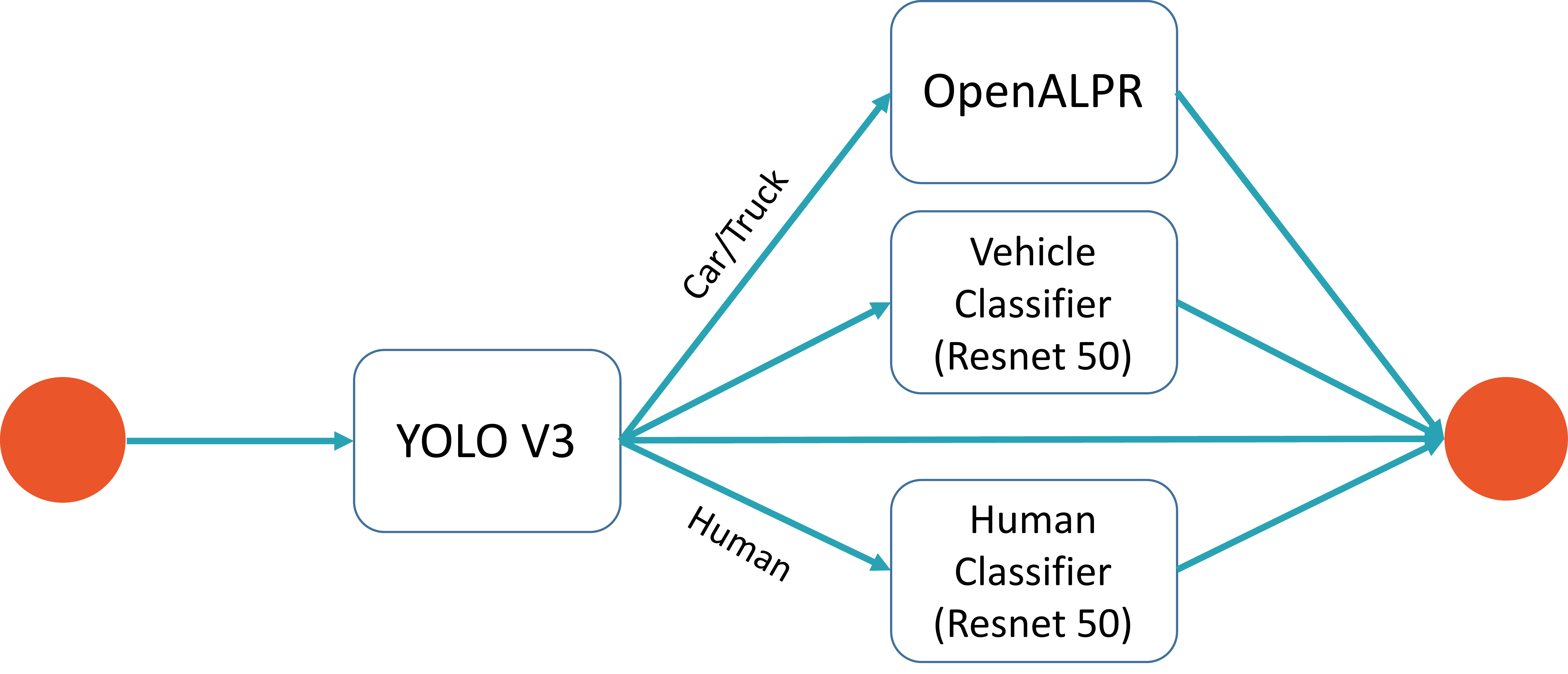}\label{fig:exp:detection}}
		\subfigure[Social Media]{\includegraphics[width=0.23\textwidth]{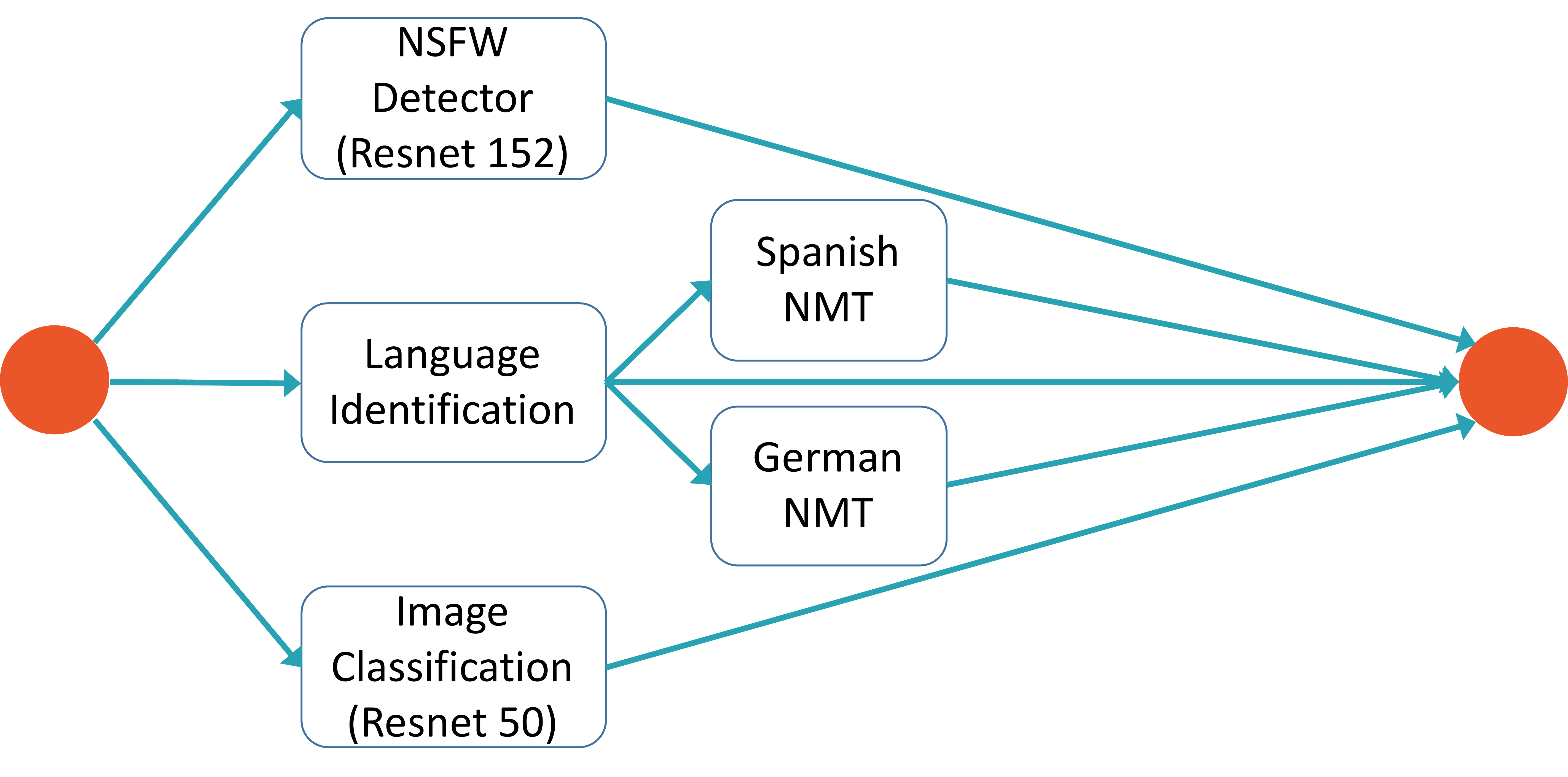}\label{fig:exp:socialmedia}}
		\subfigure[TF Cascade]{\includegraphics[width=0.23\textwidth]{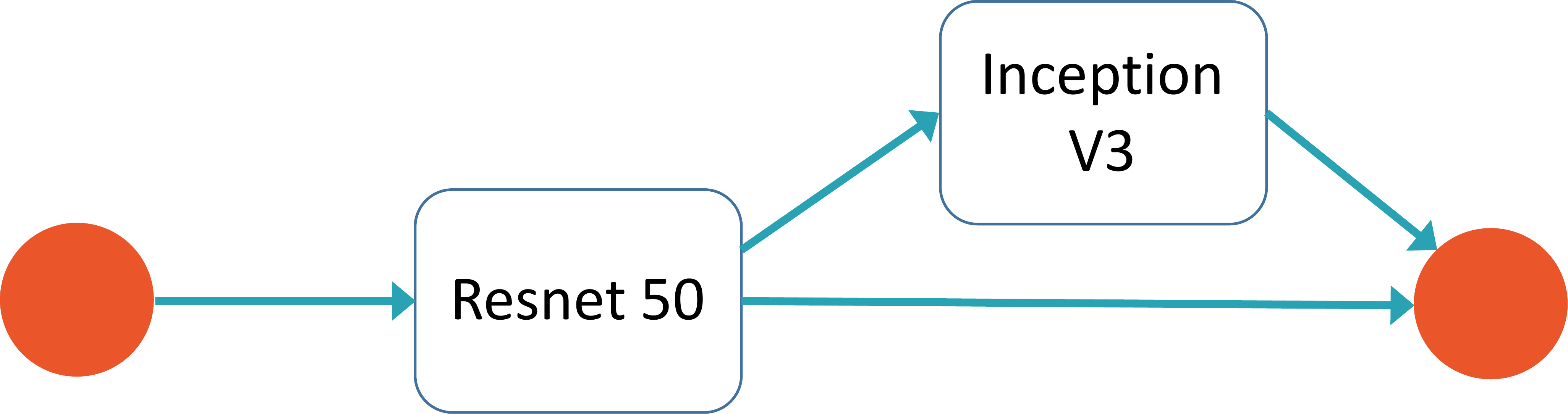}\label{fig:exp:cascade}}

	\end{center}
	\vspace{-0.2in}
	\caption{
		\footnotesize \textbf{Example Pipelines.}
			We evaluate \system on  four prediction pipelines that span a wide range of  models, control flow, and input characteristics.
	}
	\label{fig:exp:pipelines}
	\vspace{-0.15in}
\end{figure*}

Prediction pipelines combine multiple machine learning models and data transformations to support complex prediction tasks~\cite{EvansThesis}.
For instance, state-of-the-art visual question answering services~\cite{andreas15, Malinowski15} combine language models with vision models to answer the question.

A prediction pipeline can be represented as a directed acyclic graph (DAG), where each vertex corresponds to a model (e.g., mapping images to objects in the image) or a basic data transformation (e.g., extracting key frames from a video) and edges represent dataflow between these vertices.

In this paper we study several (Figure~\ref{fig:exp:pipelines}) representative prediction pipeline motifs.
The Image Processing pipeline consists of basic image pre-processing (e.g., cropping and resizing) followed by image classification using a deep neural network. 
The Video Monitoring pipeline was inspired by \cite{VideoStorm} and uses an object detection model to identify vehicles and people and then performs subsequent analysis including vehicle and person identification and license plate extraction on any relevant images.
The Social Media pipeline translates and categorizes posts based on both text and linked images by combining computer vision models with multiple stages of language models to identify the source language and translate the post if necessary.
Finally, the TensorFlow (TF) Cascade pipeline combines fast and slow TensorFlow models, invoking the slow model only when necessary.

In the Social Media, Video Monitoring, and TF Cascade pipelines, a subset of models are invoked based on the output of earlier models in the pipeline.
This conditional evaluation pattern appears in bandit algorithms~\cite{li10, Auer03} used for model personalization as well as more general cascaded prediction pipelines~\cite{McGill17, Guan17, Angelova15, Sun13}.

We show that for such pipelines \system is able to maintain latency constraints with P99 service level objectives (99\% of query latencies must be below the constraint) at low cost, even under bursty and unpredictable workloads.


\subsection{Challenges}
\label{sec:motivation:challenges}

Prediction pipelines present new challenges for the design and provisioning of prediction serving systems.
First, 
we first discuss how the proliferation of specialized hardware accelerators and the need to meet end-to-end latency constraints leads to a combinatorially large configuration space.
Second, we discuss some of the complexities of meeting tight latency SLOs under bursty stochastic query loads.
Third, we contrast this work with ideas from the data stream processing literature, which shares some \textit{structural} similarities, but is targeted at fundamentally different applications and performance goals.

\begin{figure}[tb]
	\centering
	\includegraphics[width=\columnwidth]{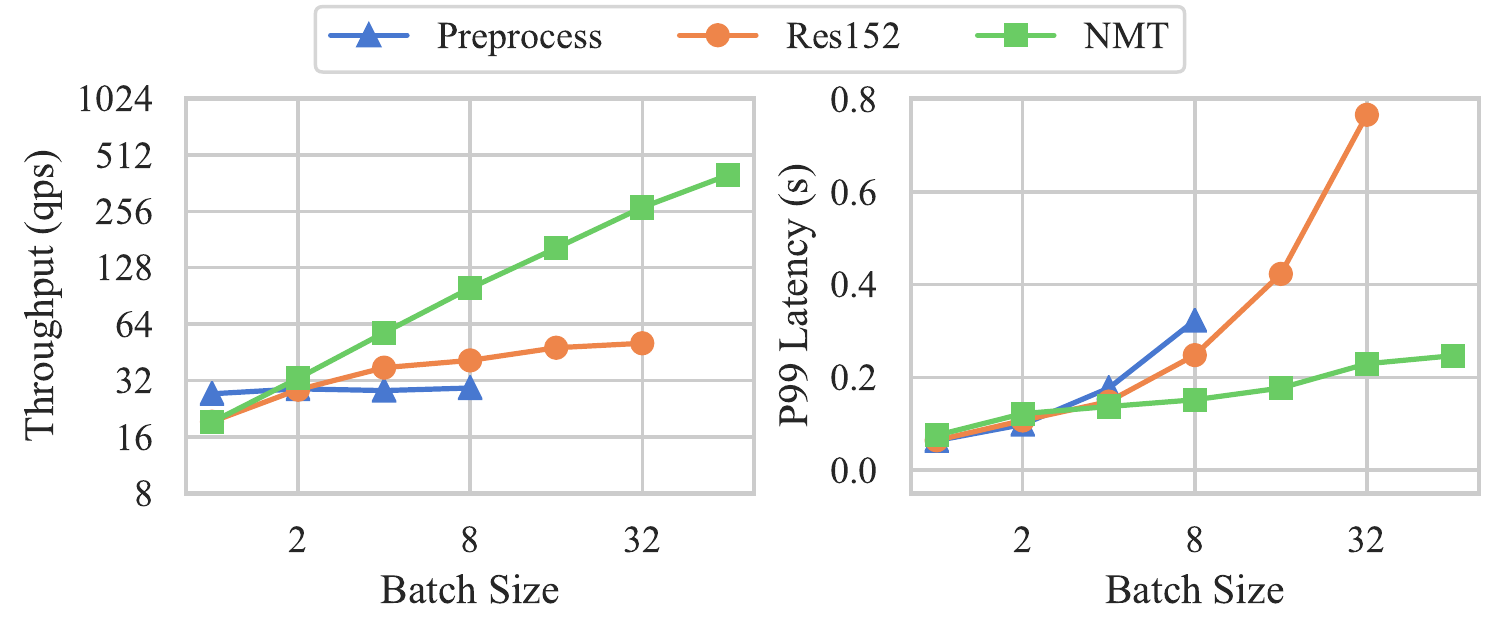}
	\vspace{-0.3in}
	\caption{
		\footnotesize \textbf{Example Model Profiles on K80 GPU.}
		The preprocess model has no internal parallelism and cannot utilize a GPU. Thus, it sees no benefit from batching. Res152 (image classification) \& TF-NMT(text translation model) benefit from batching on a GPU but at the cost of increased latency. 
	}
	\label{fig:modelprofiles}
	\vspace{-0.2in}
\end{figure}

\paragraph{Combinatorial Configuration Space}
Many machine learning models can be computationally intensive with substantial opportunities for parallelism. 
In some cases, this parallelism can result in orders of magnitude improvements in throughput and latency.
For example, in our experiments we found that TensorFlow can render predictions for the relatively large ResNet152 neural network at 0.6 queries per second (QPS) on a CPU and at 50.6 QPS on an NVIDIA Tesla K80 GPU, an 84x difference in throughput (\figref{fig:modelprofiles}).
However, not all models benefit equally from hardware accelerators.
For example, several widely used classical models (\eg decision trees~\cite{cart84}) can be difficult to parallelize on GPUs, and common data transformations (e.g. text feature extraction) often cannot be efficiently computed on GPUs.

In many cases, to fully utilize the available parallel hardware, queries must be processed in batches (\eg ResNet152 required a batch size of 32 to maximize throughput on the K80).
However, processing queries in a batch can also increase latency, as we see in~\figref{fig:modelprofiles}.  
Because most hardware accelerators operate at vector level parallelism, the first query in a batch is not returned until the last query is completed.
As a consequence, it is often necessary to set a \emph{maximum} batch size to bound query latency.
However, the choice of the maximum batch size depends on the hardware and model and affects the end-to-end latency of the pipeline.

Finally, in heavy query load settings it is often necessary to replicate individual operators in the pipeline to provide the throughput demanded by the workload.
As we scale up pipelines through replication, each operator scales differently, an effect that can be amplified by the use of conditional control flow within a pipeline causing some components to be queried more frequently than others. Low cost configurations require fine-grained scaling of each operator.

Allocating parallel hardware resources to a single model presents a complex model dependent trade-off space between cost, throughput, and latency.
This trade-off space grows exponentially with each model in a prediction pipeline.
Decisions made about the choice of hardware, batching parameters, and replication factor at one stage of the pipeline affect the set of feasible choices at the other stages due to the need to meet \emph{end-to-end} latency constraints. 
For example, trading latency for increased throughput on one model by increasing the batch size reduces the latency budget of other models in the pipeline and, as a consequence, constrains feasible hardware configurations as well.

\paragraph{Queueing Delays}
As stages of a pipeline may operate at different speeds, due to resource and model heterogeneity, it is necessary to have a queue per stage.
Queueing also allows to absorb query inter-arrival process irregularities and can be a significant end-to-end latency component. Queueing delay must be explicitly considered during pipeline configuration, as it directly depends on the relationship between the inter-arrival process and system configuration.


\paragraph{Stochastic and Unpredictable Workloads}
Prediction serving systems
must respond to bursty, stochastic query streams.
At a high-level these stochastic processes can be characterized by their average arrival rate $\lambda$ and their coefficient of variation, a dimensionless measure of variability 
defined by $CV = \frac{\sigma^2}{\mu^2}$, where $\mu = \frac{1}{\lambda}$ and $\sigma$ are the mean and standard-deviation of the query inter-arrival time.
Processes with higher CV have higher variability and often require additional over-provisioning to meet latency objectives.
Clearly, over-provisioning the whole pipeline on specialized hardware can be prohibitively expensive.  
Therefore, it is critical to be able to identify and provision the bottlenecks in a pipeline to accommodate the bursty arrival process.  
Finally, as the workload changes, we need mechanisms to monitor, quickly detect, and \textit{tune} individual stages in the pipeline.

\paragraph{Comparison to Stream Processing Systems}
Many of the challenges around configuring and scaling pipelines have been studied in the context of generic data stream processing systems~\cite{flink,samza,timely,storm}.
However, these systems focus their effort on supporting more traditional data processing workloads, which include stateful aggregation operators and support for a variety of windowing operations.
As a result, the concept of per-query latency is often ill-defined in data pipelines, and instead these systems tend to focus on maximizing throughput while avoiding backpressure, with latency as a second order performance goal (\secref{sec:prevwork}).


\section{System Design and Architecture}
\label{sec:sysarch}

In this section, we provide a high-level overview of the main system components in \system (\figref{fig:sysarch:arch}).
The system requires a \emph{planner} that operates infrequenty and re-configures the whole pipeline w.r.t. all of our control parameters and a \emph{tuner} that makes adjustments
to the pipeline configurations in response to dynamically observed query traffic patterns.

\system runs on top of any prediction serving system that
meets a few simple requirements.
The underlying serving system must be able to 1) deploy multiple replicas of a model and scale the number of replicas at runtime, 2) allow for batched inference with the ability to configure a maximum batch size, and 3) use a centralized batched queueing system to distribute batches among model replicas.
The first two properties are necessary for \system to configure the serving engine, and a centralized queueing system provides deterministic queueing behavior that can be accurately simulated by the \estimator.
In our experimental evaluation, we run \system with both Clipper~\cite{clippernsdi17} and TensorFlow Serving~\cite{tfserving}.
Both systems needed only minor modifications to meet these requirements.

\paragraph{Using \system:} 


To deploy a new prediction pipeline managed by \system, developers provide a driver program, sample query trace used for planning, and a latency service level objective.
The driver function interleaves application-specific code with asynchronous calls to models hosted in the underlying serving system to execute the pipeline.

The \planner runs as a standalone Python process that runs periodically
independent of the prediction serving framework.
The \reactive runs as a standalone process implemented in C++.
It observes the incoming arrival trace streamed to it by the centralized queueing system and triggers model addition/removal executed by serving-framework-specific APIs.


\paragraph{Low-Frequency Planning:}
The first time planning is performed, \system uses the \profiler to create performance profiles of all the individual models referenced by the driver program.
A performance profile captures model throughput as a function of hardware type and maximum batch size.
An entry in the model profile is measured empirically by evaluating the model in isolation in the given configuration using the queries in the sample trace.
The model profiles are saved and reused in subsequent runs of the planner.

The \planner finds a cost-efficient initial pipeline configuration subject to the end-to-end latency SLO and the specified arrival process.
It uses a globally-aware, cost-minimizing optimization algorithm to set the three control parameters for each model in the pipeline.
In each iteration of the optimization algorithm, the Planner uses the model profiles to select a cost-minimizing step while relying on the \estimator to check for latency constraint violations.
After the initial configuration is generated and the pipeline is deployed to serve live traffic, the Planner is re-run periodically (hours to days) on the most recent arrival history to find a cost-optimal configuration for the current workload. This also allows integrating new models and hardware.

\paragraph{High-Frequency Tuning:}
The \reactive monitors the dynamic behavior of the arrival process
to adjust per-model replication factors and maintain high SLO attainment at low cost.
The Tuner continuously monitors the current traffic envelope~\cite{netcalcbook} to detect deviations from the planning trace traffic envelope at different timescales simultaneously. 
By analyzing the timescale at which the deviation occurred, the Tuner is able to take appropriate mitigating action within seconds to ensure that SLOs are met without unnecessarily increasing cost.
It ensures that latency SLOs are maintained during unexpected changes to the arrival workload in between runs of the Planner.

\section{Low-Frequency Planning}
\label{sec:offline}

During planning, the \profiler, \estimator and \planner are used to estimate model performance characteristics and optimally provision and configure the system for a given sample workload and latency SLO.
In this section, we expand on each of these three components.

\subsection{Profiler}
\label{sec:sysarch:subsec:profiler}

The \profiler creates performance profiles for each of the models in the pipeline as a function of batch size and hardware.
Profiling begins with \system executing the sample set of queries on the pipeline.
This generates input data for profiling each of the component models individually.
We also track the frequency of queries visiting each model, called the \emph{scale factor}, $s$.
The scale factor represents the conditional probability that a model will be queried given a query entering the pipeline, independent of the behavior of any other models.
It is used by the \estimator to simulate the effects of conditional control flow on latency~(\secref{sec:sysarch:subsec:estimator}) and the \reactive to make scaling decisions (\secref{sec:reactive}).

The Profiler captures model throughput as a function of hardware type and batch size to create per-model performance profiles.
An individual model configuration corresponds to a specific value for each of these parameters as well as the model's replication factor.
Because the models scale horizontally, profiling a single replica is sufficient.
Profiling only needs to be performed once for each hardware and  batch size pair and is re-used in subsequent runs of the Planner.



\subsection{Estimator}
\label{sec:sysarch:subsec:estimator}
The \estimator is responsible for rapidly estimating the end-to-end latency of a given pipeline configuration for the sample query trace.
It takes as input a pipeline configuration, the individual model profiles, and a sample trace of the query workload, and returns accurate estimates of the latency for \textit{each query} in the trace.
The Estimator is implemented as a continuous-time, discrete-event simulator~\cite{Beck2008}, 
simulating the entire pipeline, including queueing delays.
The simulator maintains a global logical clock that is advanced from one discrete event to the next with each event triggering future events that are processed in temporal order.
Because the simulation only models discrete events, we are able to faithfully simulate hours worth of real-world traces in hundreds of milliseconds.

The Estimator simulates the deterministic behavior of queries flowing through a centralized batched queueing system.
It combines this with the model profile information which informs the simulator how long a model running on a specific hardware configuration will take to process a batch of a given size.

\subsection{Planning Algorithm}
\label{sec:planner}
\label{sec:algo:planner}

At a high-level, the planning algorithm is an iterative constrained optimization procedure that greedily minimizes cost while ensuring that the latency constraint is satisfied.
The algorithm can be divided into two phases. 
In the first (\algref{algo:planning:initialize}), it finds a feasible initial configuration that meets the latency SLO while ignoring cost. 
In the second (\algref{algo:planning:mincost}), it greedily modifies the configuration to reduce the cost while using the Estimator to identify and reject configurations that violate the latency SLO.
The algorithm converges when it can no longer make any cost reducing modifications to the configuration without violating the SLO.

\IncMargin{1em}
\begin{algorithm}[tb]
\SetKwFunction{CheckFeasible}{Feasible}
\SetKwFunction{Initialize}{Initialize}
\SetKwFunction{MaxHardware}{BestHardware}
\SetKwFunction{ServiceTime}{ServiceTime}
\SetKwFunction{ThruBottleneck}{FindMinThru}
\SetKwFunction{ConfigTemplate}{ConfigTemplate}
\SetKwProg{Fn}{Function}{:}{}
\SetKwData{Config}{Config}
\SetKwData{EmptyConfig}{EmptyConfig}
\SetKwData{True}{True}
\SetKwData{False}{False}
\Fn{\Initialize{pipeline, slo}}{
    \ForEach{model \textnormal{in} pipeline}{
        $model$.batchsize = 1\;
        $model$.replicas = 1\;
        $model$.hw = \MaxHardware{model}\;  
    }
    \If{\ServiceTime{pipeline} $\leq$ slo}{
        \KwRet{\False}\;
    }
    \Else{
        \While{\textnormal{not} \CheckFeasible{pipeline, slo}}{
            $model$ = \ThruBottleneck{pipeline}\;
            $model$.replicas += 1\;
        }
        \KwRet{pipeline}\;
    }
}
\caption{\footnotesize Find an initial, feasible configuration}\label{algo:planning:initialize}
\end{algorithm}
\DecMargin{1em}

\paragraph{Initialization (\algref{algo:planning:initialize}):}
First, an initial latency-minimizing configuration is constructed by setting the batch size to 1 using the lowest latency hardware available for each model (lines 2-5).
If the service time under this configuration (the sum of the processing latencies of all the models on the longest path through the pipeline DAG) is greater than the SLO then the latency constraint is infeasible given the available hardware and the Planner terminates (lines 6-7).
Otherwise, the Planner then iteratively determines the \emph{throughput bottleneck} and increases that model's replication factor until it is no longer the bottleneck (lines 9-11).


\IncMargin{1em}
\begin{algorithm}[tb]
\SetKwFunction{CheckFeasible}{Feasible}
\SetKwFunction{Initialize}{Initialize}
\SetKwFunction{MinimizeCost}{MinimizeCost}
\SetKwFunction{MaxHardware}{MaxHardware}
\SetKwFunction{ServiceTime}{ServiceTime}
\SetKwFunction{CostBottleneck}{FindMaxCost}
\SetKwFunction{IncreaseBatch}{IncreaseBatch}
\SetKwFunction{RemoveReplica}{RemoveReplica}
\SetKwFunction{DowngradeHW}{DowngradeHW}
\SetKwProg{Fn}{Function}{:}{}
\SetKwData{True}{True}
\SetKwData{False}{False}
\Fn{\MinimizeCost{pipeline, slo}}{
    $pipeline$ = \Initialize{pipeline, slo}\;
    \If{pipeline == \False}{
        \Return{\False}\;
    }
    $actions$ = [\IncreaseBatch, \RemoveReplica, \DowngradeHW]\;
    \Repeat{best == \textnormal{\texttt{NULL}}}{
        $best$ = \texttt{NULL}\;
        \ForEach{model \textnormal{in} pipeline}{
            \ForEach{action \textnormal{in} actions}{
                $new$ = $action$($model$, $pipeline$)\;
                \If{\CheckFeasible{$new$}}{
                    \If{$new$.cost < $best$.cost}{
                        $best = new$\;
                    }
                }
            }
        }
        \If{best \textnormal{is not \texttt{NULL}}}{
            $pipeline$ = $best$\;
        }
    }
    \Return{pipeline}\;
}
\caption{\footnotesize Find the min-cost configuration}\label{algo:planning:mincost}
\end{algorithm}
\DecMargin{1em}

\paragraph{Cost-Minimization (\algref{algo:planning:mincost}):}
In each iteration of the cost-minimizing process, the Planner considers three candidate modifications for each model: increase the batch size, decrease the replication factor, or downgrade the hardware (line 5), searching for the modification that maximally decreases cost while still meeting the latency SLO.
It evaluates each modification on each model in the pipeline (lines 8-10), discarding candidates that violate the latency SLO according to the Estimator (line 11).

The \textbf{batch size} only affects throughput and does not affect cost.
It will therefore only be the cost-minimizing modification if the other two would create infeasible configurations.
Increasing the batch size does increase latency.
The batch size is increased by factors of two as the throughput improvements from larger batch sizes have diminishing returns (observe ~\figref{fig:modelprofiles}).
In contrast, decreasing the \textbf{replication factor} directly reduces cost.
Removing replicas is feasible when a previous iteration of the algorithm has increased the batch size for a model, increasing the per-replica throughput.



\textbf{Downgrading hardware} is more involved than the other two actions, as the batch size and replication factor for the model must be re-evaluated to account for the differing batching behavior of the new hardware. 
It is often necessary to reduce the batch size and increase replication factor to find a feasible pipeline configuration.
However, the reduction in hardware price sometimes compensates for the increased replication factor.
For example, in~\figref{fig:opt-sensitivity}, the steep decrease in cost when moving from an SLO of 0.1 to 0.15 can be largely attributed to downgrading the hardware of a language identification model from a GPU to a CPU.

To evaluate a hardware downgrade, we first freeze the configurations of the other models in the pipeline and perform the initialization stage for that model using the next cheapest hardware. 
The planner then performs a localized version of the cost-minimizing algorithm to find the batch size and replication factor for the model on the newly downgraded resource allocation needed to reduce the cost of the previous configuration.
If there is no cost reducing feasible configuration the hardware downgrade action is rejected.


At the point of termination, the planning algorithm provides the following guarantees:
(1) If there is a configuration that meets the latency SLO, then the algorithm will return a valid configuration.
(2) There is no single action that can be taken to reduce cost without violating the SLO.

\section{High-Frequency Tuning}
\label{sec:online}
\label{sec:reactive}
\label{sec:algo:reactive}

\system's Planner finds an efficient, low-cost configuration that is guaranteed to meet the provided latency objective.
However, this guarantee only holds for the sample planning workload provided to the planner.
Real workloads evolve over time, changing in both arrival rate (change in $\lambda$) as well as becoming more or less bursty (change in CV).
When the serving workload deviates from the sample, the original configuration will either suffer from queue buildups leading to SLO misses or be over-provisioned and incur unnecessary costs.
The \reactive both \emph{detects} these changes as they occur and takes the appropriate \emph{scaling action} to maintain both the latency constraint and cost-efficiency objective.

In order to maintain P99 latency SLOs, the Tuner must be able to detect changes in the arrival workload dynamics across multiple timescales simultaneously.
The Planner guarantees that the pipeline is adequately provisioned for the sample trace.
The Tuner's detection mechanism detects when the current request workload exceeds the sample workload.
To do this, we draw on the idea of traffic envelopes from network calculus~\cite{netcalcbook} to characterize the workloads.

\begin{figure}[tb]
	\centering
	\includegraphics[width=0.9\columnwidth]{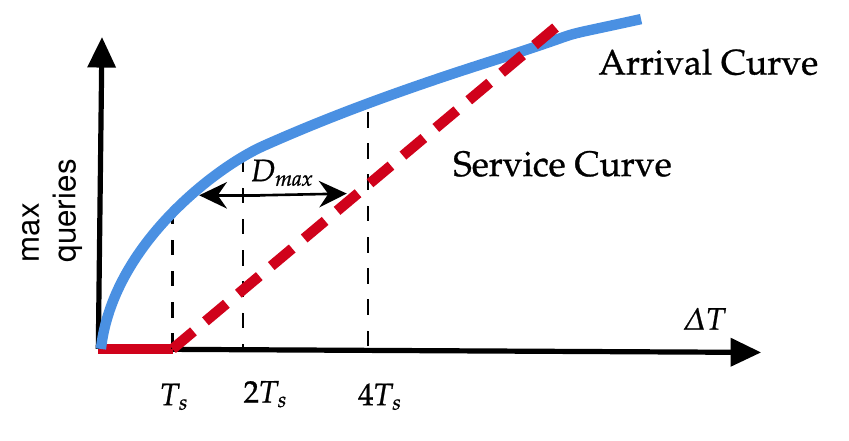}
	\label{fig:algo:reactive}
	\caption{\textbf{Arrival and Service Curves.} The arrival curve captures the maximum number of queries to be expected in any interval of time $x$ seconds wide. The service curve plots the expected number of queries processed in an interval of time $x$ seconds wide.}
    \vspace{-0.2in}
\end{figure}
A traffic envelope for a workload is constructed by sliding a window of size
$\Delta T_i$ over the workload's inter-arrival process and capturing the maximum number of queries seen anywhere within this window. Thus, each $x=\Delta T_i$ is mapped to $y=q_i$ (number of queries) for all $x$ over the duration of a trace. This powerful characterization captures how much the workload can burst in any given interval of time. In practice, we discretize the x-axis by setting the smallest $\Delta T_i$ to $T_s$, the service time of the system, and then double the window size up to 60 seconds. For each such interval, the maximum arrival rate $r_i$ for this interval can be computed as $r_i = \frac{q_i}{\Delta T_i}$. By measuring $r_i$ across all $\Delta T_i$ \textit{simultaneously} we capture a fine-grain characterization of the arrival workload that enables simultaneous detection of changes in both short term (burstiness) and long term (average arrival rate) traffic behavior.

\paragraph{Initialization}
During planning, the Planner constructs the traffic envelope for the sample arrival trace.
The Planner also computes the max-provisioning ratio for each model $\rho_m = \frac{\lambda}{\mu_m}$, the ratio of the arrival rate $\lambda$ to the maximum throughput of the model $\mu$ in its current configuration.
While the max-provisioning ratio is not a fundamental property of the pipeline, it provides a useful heuristic to measure how much ``slack'' the Planner has determined is needed for this model to be able to absorb bursts and still meet the SLO.
The Planner then provides the Tuner with the traffic envelope for the sample trace, the max-provisioning ratio $\rho_m$ and single replica throughput $\mu_m$ for each model in the pipeline.

In the low-latency applications that \system targets, failing to scale up the pipeline in the case of an increased workload results in missed latency objectives and degraded quality of service, while failing to scale down the pipeline in the case of decreased workload only results in slightly higher costs.
We therefore handle the two situations separately.

\paragraph{Scaling Up}
The Tuner continuously computes the traffic envelope for the current arrival workload.
This yields a set of arrival rates for the current workload that can be directly compared to those of the sample workload.
If any of the current rates exceed their corresponding sample rates, the pipeline is underprovisioned and the Tuner checks whether it add replicas for any models in the pipeline.

At this point, not only has the Tuner detected that rescaling may be necessary, it also knows what arrival rate it needs to reprovision the pipeline for: the current workload rate $r_{max}$ that triggered rescaling.
If the overall $\lambda$ of the workload has not changed but it has become burstier, this will be a rate computed with a smaller $\Delta T_i$, and if the burstiness of the workload is stationary but the $\lambda$ has increased, this will be a rate with a larger $\Delta T_i$.
In the case that multiple rates have exceeded their sample trace counterpart, we take the max rate.

To determine how to reprovision the pipeline,
the Tuner computes the number of replicas needed for each model to process $r_{max}$ as
$k_m = \ceil{\frac{r_{max} s_m}{\mu_m \rho_m}}$.
$s_m$ is the scale factor for model $m$, which prevents over-provisioning for a model that only receives a portion of the queries due to conditional logic.
$\rho_m$ is the max-provisioning ratio, which ensures enough slack remains in the model to handle bursts.
The Tuner then adds the additional replicas needed for any models that are detected to be underprovisioned.

\paragraph{Scaling Down}
\system takes a conservative approach to scaling down the pipeline to prevent unnecessary configuration oscillation which can cause SLO misses.
Drawing on the work in~\cite{autoscale}, the Tuner waits for a period of time after any configuration changes to allow the system to stabilize before considering any down scaling actions.
\system uses a delay of 15 seconds (3x the 5 second activation time of spinning up new replicas in the underlying prediction serving frameworks), but the precise value is unimportant as long as it provides enough time for the pipeline to stabilize after a scaling action.
Once this delay has elapsed, the Tuner computes the max request rate $\lambda_{new}$ that has been observed over the last 30 seconds, using 5 second windows.

The Tuner computes the number of replicas needed for each model to process $\lambda_{new}$ similarly to the procedure for scaling up, setting $k_m = \ceil{\frac{\lambda_{new} s_m}{\mu_m \rho_{p}}}$.
In contrast to scaling up,
when scaling down we use the minimum max provisioning factor in the pipeline $\rho_{p} = \min{\left(\rho_m \forall m \in \textnormal{models}\right)}$.
Because the max provisioning factor is a heuristic that has some dependence on the sample trace, using the min across the pipeline provides a more conservative downscaling algorithm and ensures the Tuner is not overly aggressive in removing replicas.
If the workload has dropped substantially, the next time the Planner runs it will find a new lower-cost configuration that is optimal for the new workload.

\section{Experimental Setup}
\label{sec:expsetup}

To evaluate \system{} we constructed four prediction pipelines (\figref{fig:exp:pipelines}) representing common application domains and using models trained in a variety of machine learning frameworks~\cite{pytorch,tensorflow,yolo,openalpr}.
We configure each pipeline with varying input arrival processes and latency budgets.
We evaluate the latency SLO attainment and pipeline cost under a range of both synthetic and real world workload traces.
\paragraph{Coarse-Grained Baseline Comparison}
\label{sec:syscomp}

Current prediction serving systems do not provide
functionality for provisioning and managing prediction pipelines with end-to-end latency constraints.
Instead, the individual pipeline components are each deployed as a separate microservice to a prediction serving system such as~\cite{tfserving,sagemaker,clippernsdi17,tensorrtserver} and a pipeline is manually constructed by individual calls to each service.


Any performance tuning for end-to-end latency or cost treats the entire pipeline as a single black-box service and tunes it as a whole.
We therefore use this same approach as our baseline for comparison.
Throughout the experimental evaluation we refer to this as the \emph{Coarse-Grained} baseline.
We deploy pipelines configured with both \system and the coarse-grained baseline to the same underlying prediction-serving franework.
All experiments used Clipper~\cite{clippernsdi17} as the prediction-serving framework except for those in~\figref{fig:eval:gen} which compare \system running on Clipper and TensorFlow Serving~\cite{tfserving}.
Both prediction-serving frameworks were modified to add a centralized batched queueing system.


We use the techniques proposed in~\cite{autoscale} to do both low-frequency planning and high-frequency tuning for the coarse-grained pipelines as a baseline for comparison.
In this baseline, we profile the entire pipeline as a single black box to identify the single maximum batch size capable of meeting the SLO, in contrast to \system's per-model profiling.
The pipeline is then replicated as a single unit to achieve the required throughput as measured on the same sample arrival trace used by the \planner.
We evaluate two strategies for determining required throughput.
\emph{CG-Mean} uses the mean request rate computed over the arrival trace while \emph{CG-Peak} determines the peak request rate in the trace computed using a sliding window of size equal to the SLO.
The coarse-grained tuning mechanism scales the number of pipeline replicas using the scaling algorithm introduced in~\cite{autoscale}.

\paragraph{Physical Execution Environment}
\label{sec:expsetup:phys}

We ran all experiments in a distributed cluster on Amazon EC2. The pipeline driver client was deployed on an \texttt{m4.16xlarge} instance which has 64 vCPUs, 256 GiB of memory, and 25Gbps networking across two NUMA zones. 
We used large client instance types to ensure that network bandwidth from the client is not a bottleneck.
Models were deployed to a cluster of up to 16 \texttt{p2.8xlarge} GPU instances. 
This instance type has 8 NVIDIA K80 GPUs, 32 vCPUs, 488.0 GiB of memory and 10Gbps networking all within a single NUMA zone.
All instances ran Ubuntu 16.04 with Linux Kernel version 4.4.0.

CPU costs were computed by dividing the total hourly cost of an instance by the number of CPUs.
GPU costs were computed by taking the difference between a GPU instance and its equivalent non-GPU instance (all other hardware matches), then dividing by the number of GPUs.
This cost model provides consistent prices across instance sizes.

\paragraph{Workload Setup}

We generated synthetic traces by sampling inter-arrival times from a gamma distribution with differing mean $\mu$ to vary the request rate, and coefficient of variation CV to vary the workload burstiness. 
When reporting performance on a specific workload as characterized by $\lambda = \frac{1}{\mu}$ and CV, a trace for that workload was generated once and reused across all comparison points to provide a more direct comparison of performance.
We generated separate traces with the same performance characteristics for profiling and evaluation to avoid overfitting to the sample trace.

To generate synthetic time-varying workloads, we evolve the workload generating function between different Gamma distributions over a specified period of time, the transition time. This allows us to generate workloads that vary in mean throughput, CV, or both, and thus evaluate the performance of the \reactive under a wide range of conditions.

In~\figref{fig:autoscale} we evaluate \system{} on traces derived from real workloads studied in the AutoScale system~\cite{autoscale}.
These workloads only report the average request rate each minute for an hour, rather than providing the full trace of query inter-arrival times.
To derive traces from these workloads, we followed the approach used by~\cite{autoscale} to re-scale the max throughput to 300 QPS, the maximum throughput supported by the coarse-grained baseline pipelines on a 16 node (128 GPU) cluster.
We then iterated through each of the mean request rates in the workload and sample from a Gamma distribution with CV 1.0 for 30 seconds. 
We use the first 25\% of the trace as the sample for the Planner, and the remaining 75\% as the live serving workload (see~\figref{fig:autoscale}).

\section{Experimental Evaluation}
\label{sec:eval}

In this section we evaluate \system's performance.
First, we evaluate end-to-end performance of \system relative to current state of the art methods for configuring and provisioning prediction pipelines with end-to-end latency constraints (\secref{sec:eval:e2e}).
We show that \system outperforms the baselines on latency SLO attainment and cost for  synthetic and real-world derived workloads with both stable and unpredictable workload dynamics.
Second, we 
demonstrate that \system is robust to unplanned dynamics of the arrival process~(\secref{sec:eval:sensitivity}): changes in the arrival rate as well as unexpected inter-arrival bursts, as the \reactive rapidly re-scales the pipeline in response to these changes.
Third, we perform an ablation study to show that the system benefits from both
the low-frequency planning and high-frequency tuning.
We conclude by showing that \system composes with multiple underlying prediction-serving frameworks~(\secref{sec:eval:gen}).

\subsection{End-to-end Evaluation}
\label{sec:eval:e2e}
We first establish that \system's planning and tuning components
outperform state-of-the-art pipeline-level configuration alternatives in an end-to-end evaluation (\secref{sec:eval:e2e:reactive}).
\system is able to achieve the same throughput at significantly lower cost,
while maintaining zero or near-zero latency SLO miss rate.

\begin{figure*}[th]
\centering
\subfigure[Image Processing CV 1.0]{
    \includegraphics[width=0.23\textwidth]{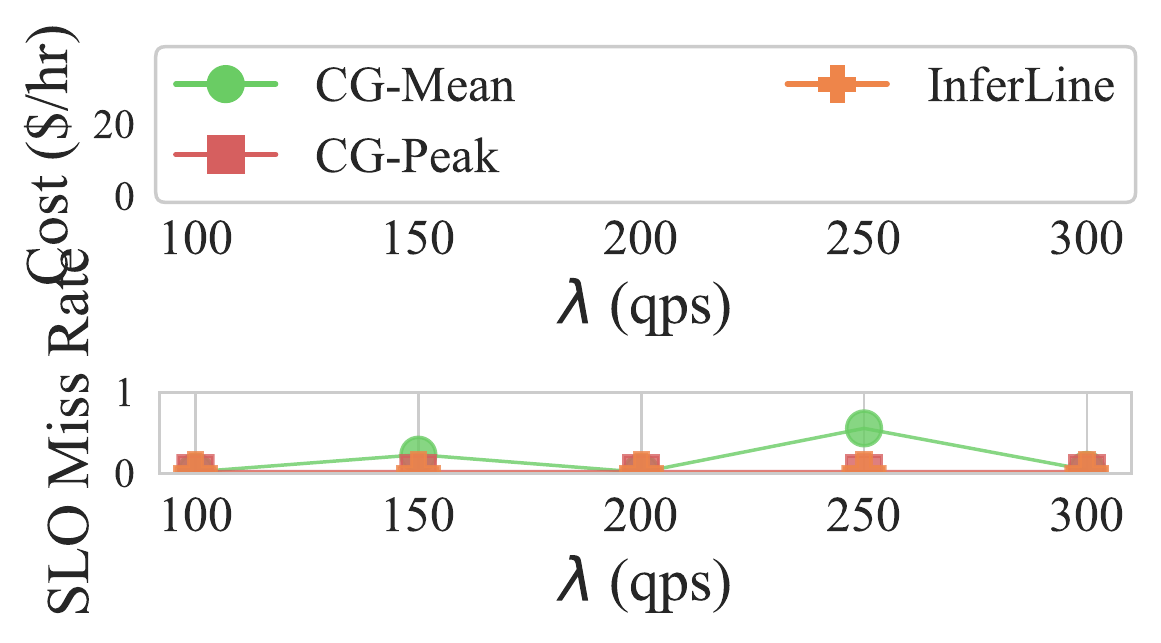}
    \label{fig:exp:proactive:heavycpu:slo15cv1}
}
\subfigure[Image Processing CV 4.0]{
    \includegraphics[width=0.23\textwidth]{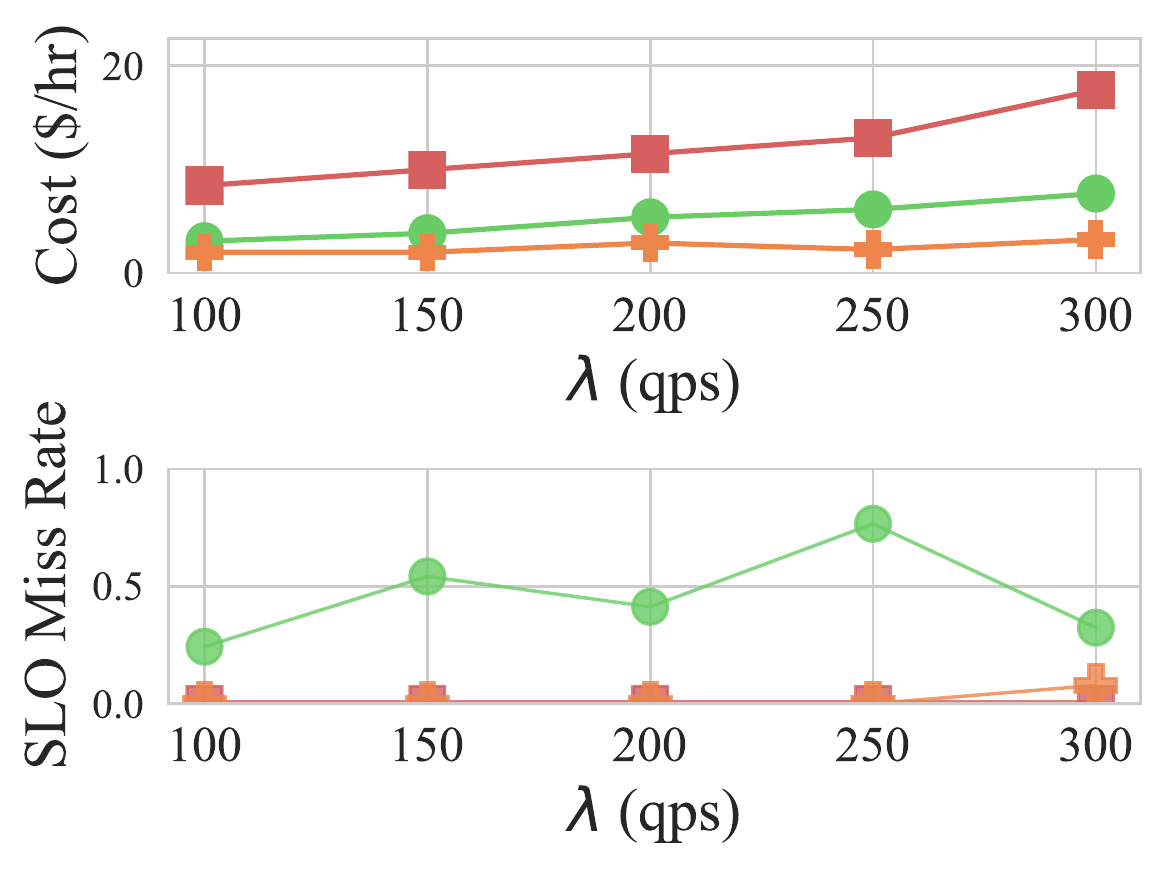}
    \label{fig:exp:proactive:heavycpu:slo15cv4}
}
\subfigure[Video Monitoring CV 1.0]{
    \includegraphics[width=0.23\textwidth]{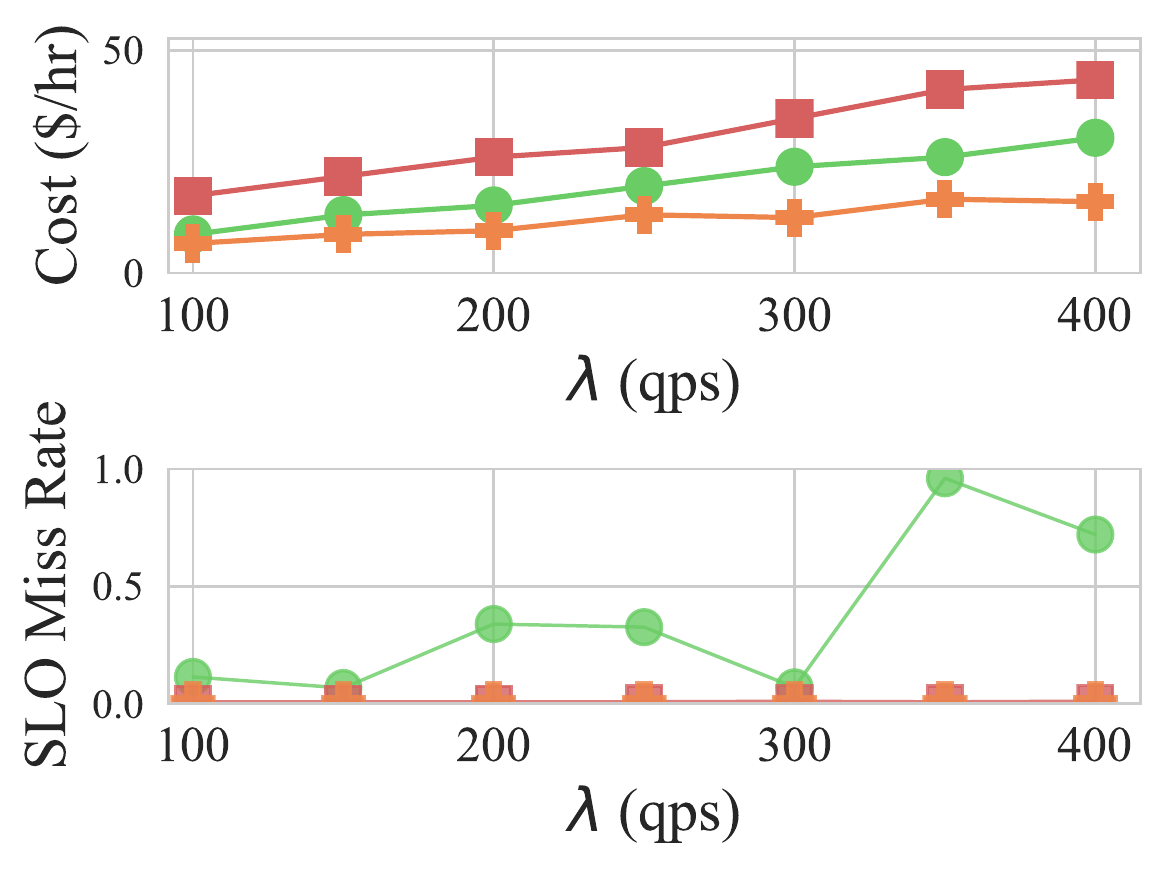}
    \label{fig:exp:proactive:detection:slo15cv1}
}
\subfigure[Video Monitoring CV 4.0]{
    \includegraphics[width=0.23\textwidth]{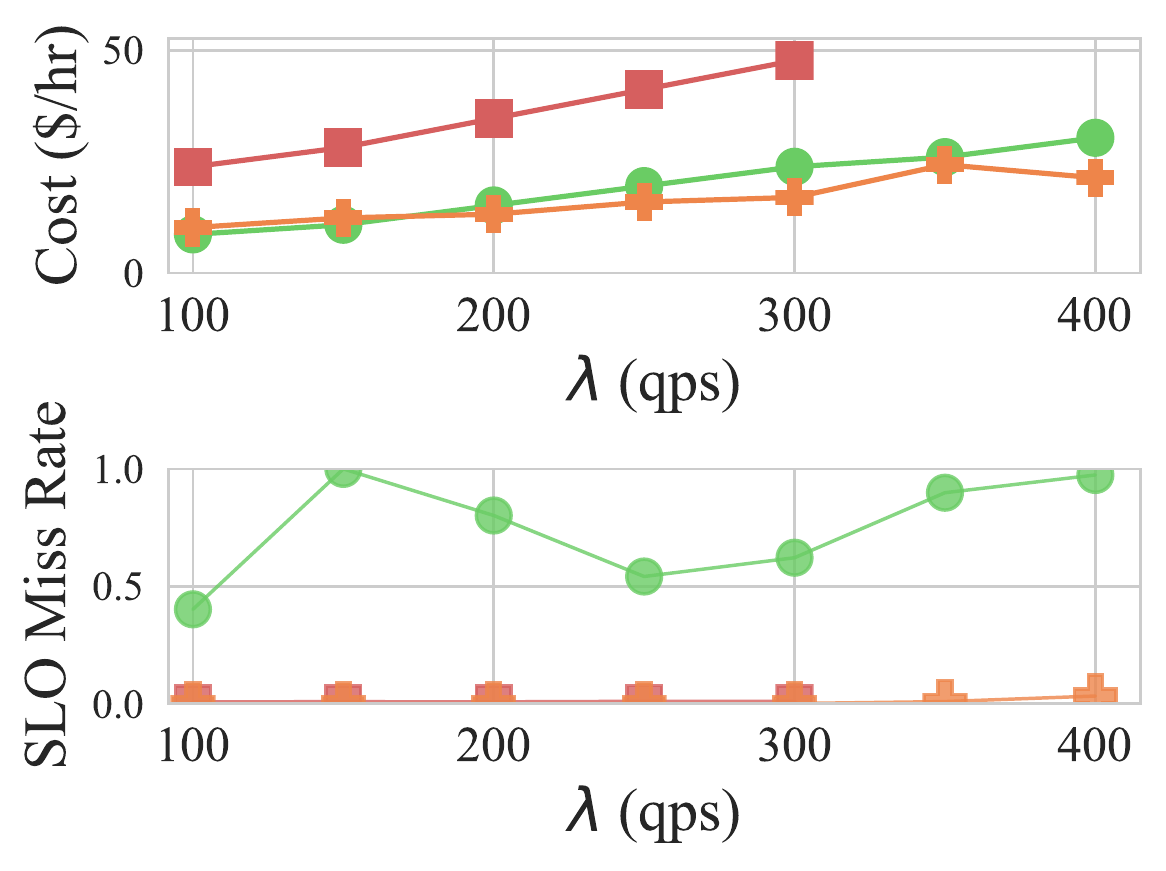}
    \label{fig:exp:proactive:detection:slo15cv4}
}
\vspace{-0.15in}
\caption[short]{\footnotesize \textbf{Comparison of \system's Planner to coarse-grained baselines (150ms SLO)} \system outperforms both baselines, consistently providing both the lowest cost configuration and highest SLO attainment (lowest miss rate).
CG-Peak was not evaluated on $\lambda$ > 300 because the configurations exceeded cluster capacity.}
\vspace{-0.2in}
\label{fig:exp:proactive}
\end{figure*}

\paragraph{Low-Frequency Planning}
\label{sec:eval:e2e:proactive}
In the absence of a workload-aware planner~(\secref{sec:planner}),
the options are limited to either 
(a) provisioning for the peak (CG Peak), or 
(b) provisioning for the mean (CG Mean) request rate. We compare
\system to these two end-points of the configuration continuum across 2 pipelines~(\figref{fig:exp:proactive}).
\system meets latency SLOs at the lowest cost. CG Peak meets SLOs, but at much higher cost,
particularly for burstier workloads.
And CG Mean is not provisioned to handle arrival bursts which results in high SLO miss rates.

The \planner consistently finds lower cost configurations than both coarse-grained provisioning strategies and is able to achieve up to a \emph{7.6x reduction in cost} by minimizing pipeline imbalance.
Finally, we observe that the Planner consistently finds configurations that meet the SLO for workloads with the same characteristics as the sample trace used for planning.
Next, we evaluate the Tuner's ability to meet SLOs during \emph{unexpected} changes in workload.



\paragraph{High-Frequency Tuning}
\label{sec:eval:e2e:reactive}

\begin{figure*}[th]
\centering
\subfigure[AutoScale Big Spike]{
    \centering
    \includegraphics[width=\columnwidth]{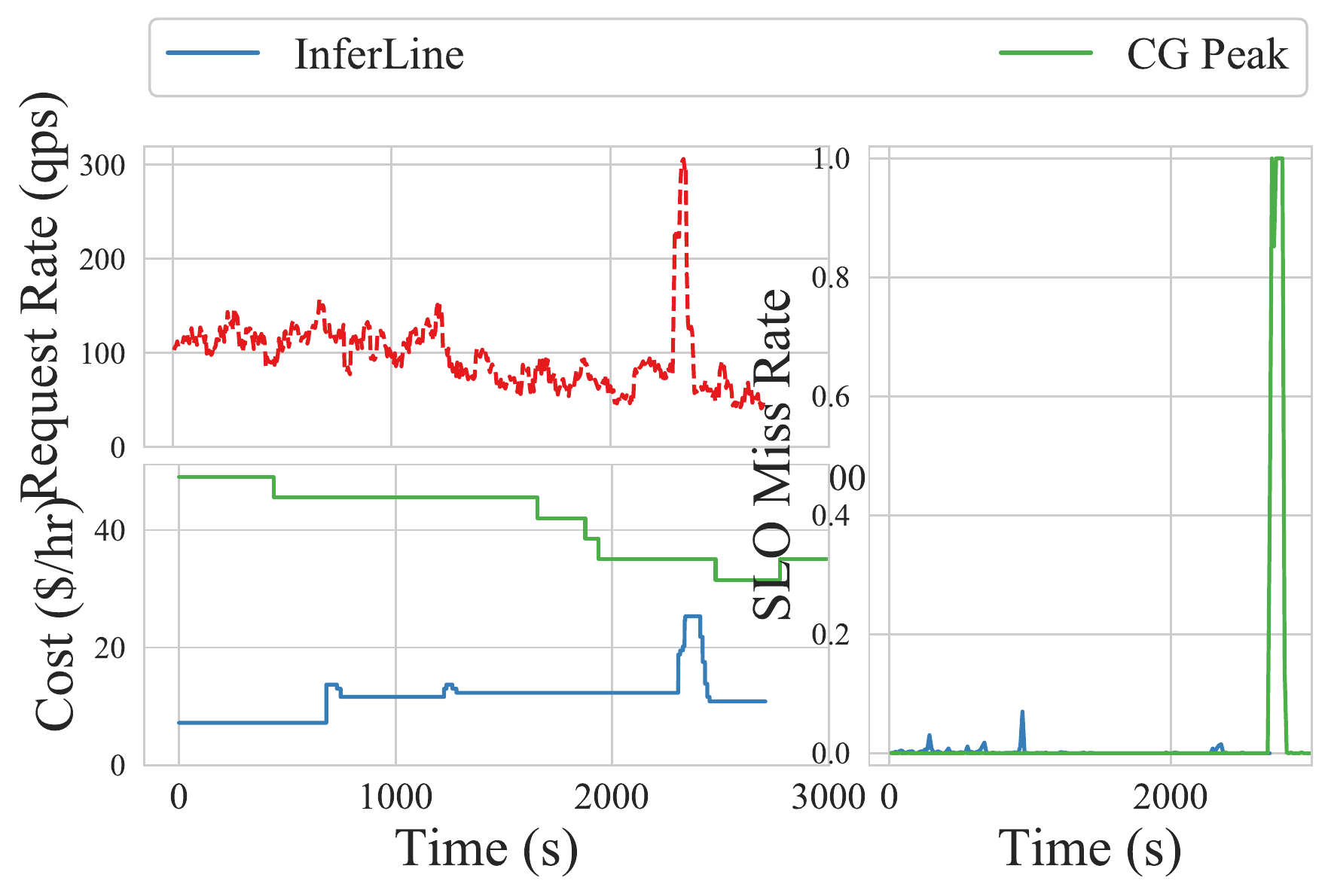}
    \label{fig:autoscale:bigspike}
}
\subfigure[AutoScale Slowly Varying]{
    \centering
    \includegraphics[width=\columnwidth]{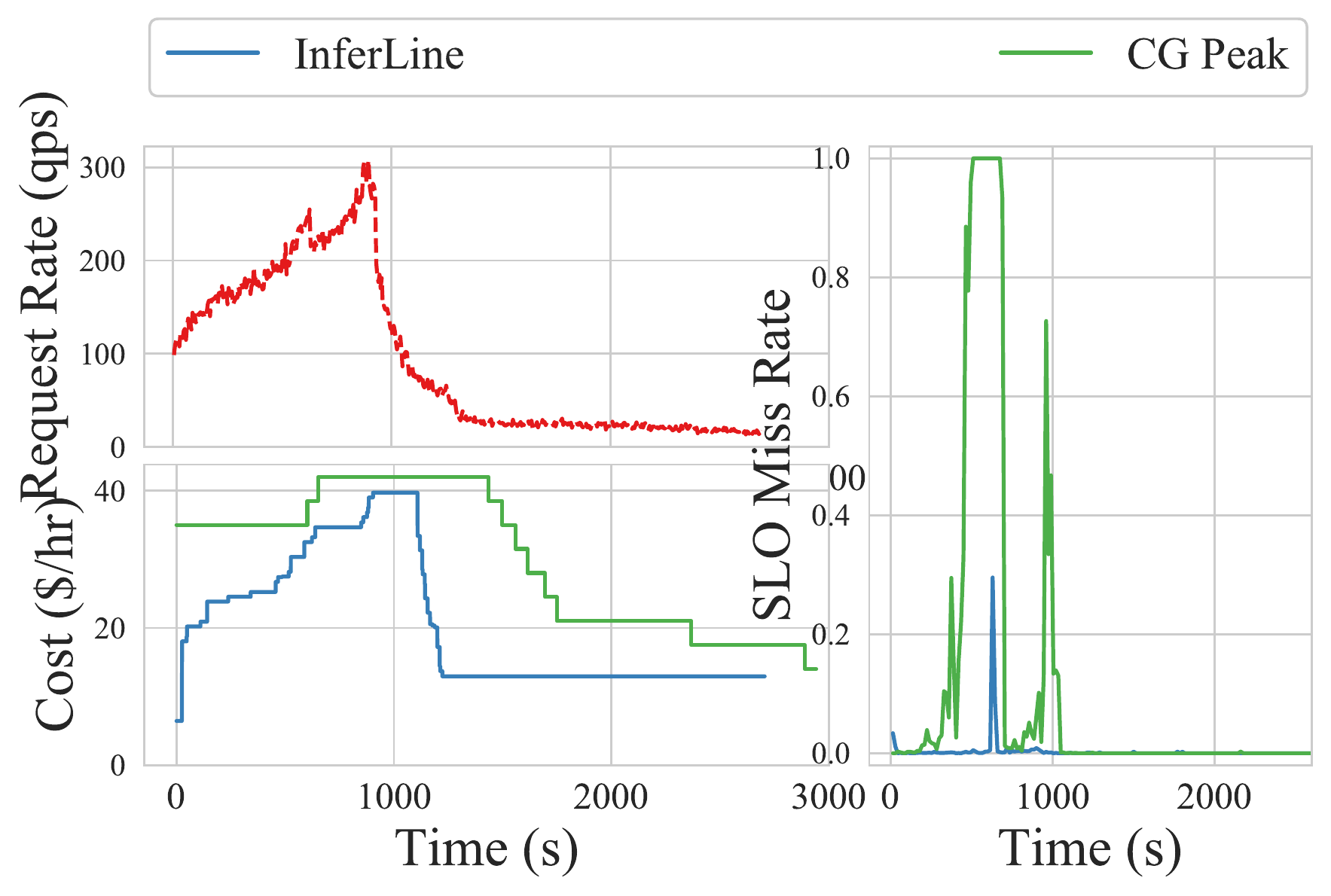}
    \label{fig:autoscale:slowlyvarying}
}
\vspace{-0.15in}
\caption[short]{\footnotesize\textbf{Performance comparison of the high-frequency tuning algorithms on traces derived from real workloads}. These are the same workloads evaluated in~\cite{autoscale} which forms the basis for the coarse-grained baseline. Both workloads were evaluated on the Social Media pipeline with a 150ms SLO. In~\figref{fig:autoscale:bigspike}, \system maintains a 99.8\% SLO attainment overall at a total cost of \$8.50, while the coarse-grained baseline has a 93.7\% SLO attainment at a cost of \$36.30.
In~\figref{fig:autoscale:slowlyvarying}, \system has a 99.3\% SLO attainment at a cost of \$15.27, while the coarse-grained baseline has a 75.8\% SLO attainment at a cost of \$24.63, a 34.5x lower SLO miss rate.}
\vspace{-0.15in}
\label{fig:autoscale}
\end{figure*}

\system is able to 
(1) maintain a negligible SLO miss rate, and
(2) and reduce cost by up to 4.2x when compared to the state-of-the-art approach~\cite{autoscale}
when handling unexpected changes in the arrival rate and burstiness.
In \figref{fig:autoscale} we evaluate the \emph{Social Media} pipeline on 2 traces derived from real workloads studied in~\cite{autoscale}.
The \planner finds a \emph{5x cheaper} initial configuration than coarse-grained provisioning~(\figref{fig:autoscale:bigspike}).
Both systems achieve near-zero SLO miss rates throughout most of the workload, and when the big spike occurs we observe that \system's \reactive quickly reacts by scaling up the pipeline as described in~\secref{sec:algo:reactive}.
As soon as the spike dissipates, \system scales the pipeline down to maintain a cost-efficient configuration.
In contrast, the coarse-grained tuning mechanism operates much slower and, therefore, is ill-suited for reacting to rapid changes in the request rate of the arrival process.

In \figref{fig:autoscale:slowlyvarying}, \system scales up the pipeline smoothly and recovers rapidly from an instantaneous spike, unlike the CG baseline.
As the workload drops quickly after 1000 seconds, \system rapidly responds by shutting down replicas to reduce cluster cost.
In the end, \system and the coarse-grained pipelines converge to similar costs due to the low terminal request rate which hides the effects of pipeline imbalance, but \system has a \emph{34.5x lower SLO miss rate} than the baseline.


\begin{figure*}[th]
\centering
\subfigure[Image Processing]{
    \centering
    \includegraphics[width=0.47\textwidth]{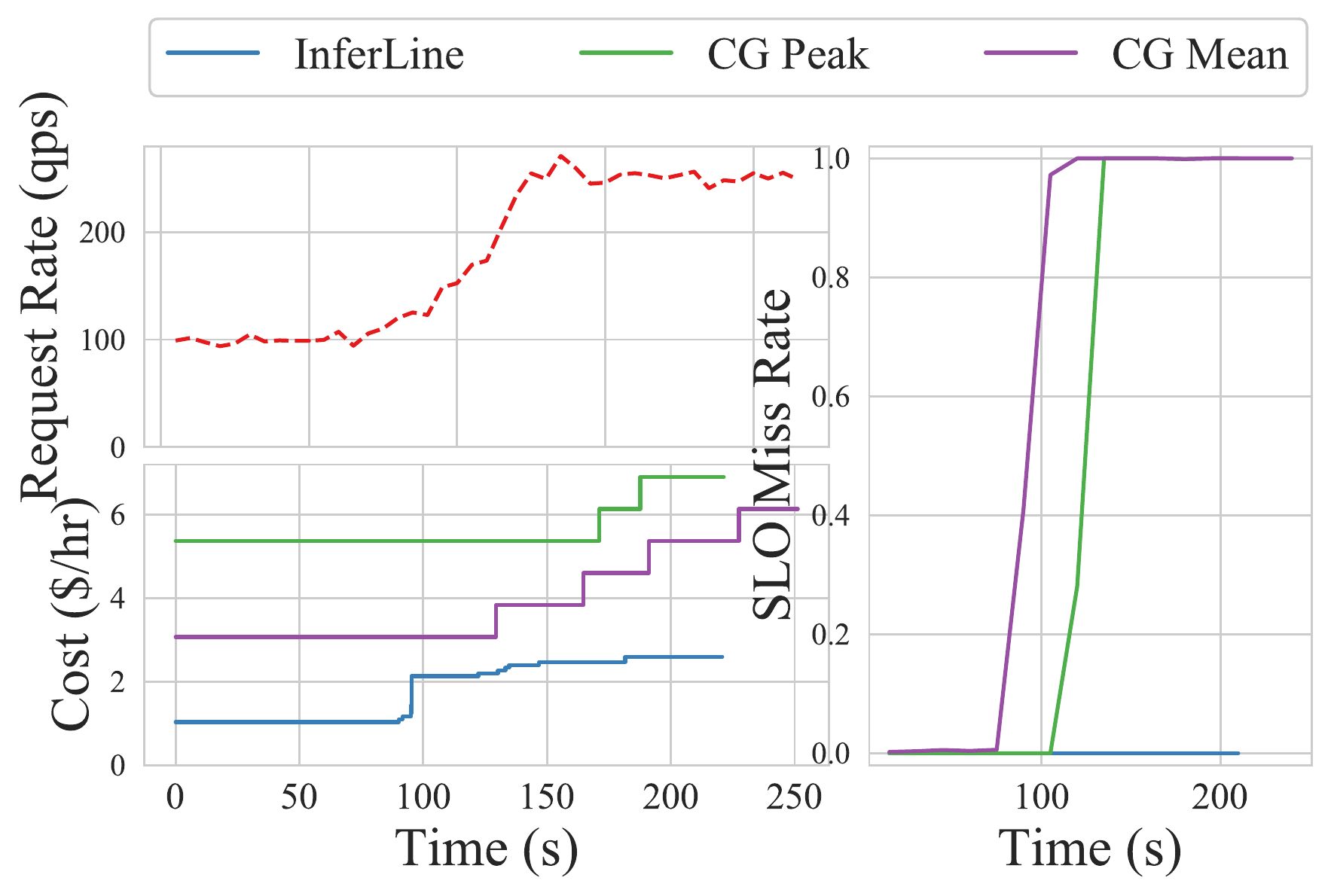}
}
\subfigure[Social Media]{
    \centering
    \includegraphics[width=0.47\textwidth]{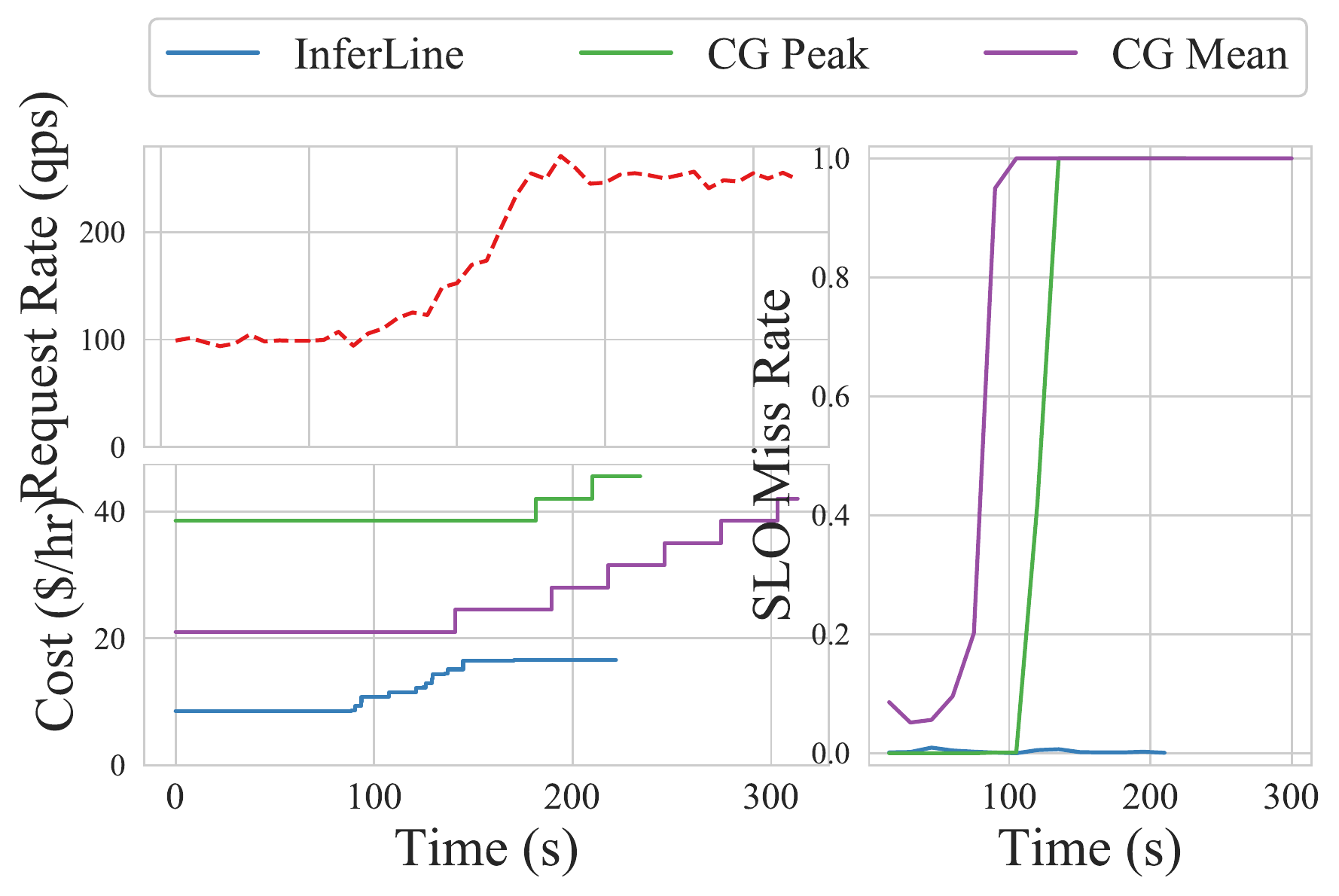}
}
\vspace{-0.15in}
\caption{\footnotesize\textbf{Performance comparison of the high-frequency tuning algorithms on synthetic traces with increasing arrival rates.} We observe that \system outperforms both coarse-grained baselines on cost while maintaining a near-zero SLO miss rate for the entire duration of the trace.}
\vspace{-0.1in}
\label{fig:exp:reactivesynthetic}
\end{figure*}

We further evaluate the differences between the \system and coarse-grained tuning algorithms on a set of synthetic workloads with increasing arrival rates in~\figref{fig:exp:reactivesynthetic}.
We observe that the traffic envelope monitoring described in~\secref{sec:reactive} enables \system to detect the increase in arrival rate earlier and therefore scale up the pipeline sooner to maintain a low SLO miss rate.
In contrast, the coarse-grained baseline only reacts to the increase in request rate at the point when the pipeline is overloaded and therefore reacts when the pipeline is already in an infeasible configuration.
The effect of this delayed reaction is compounded by the longer provisioning time needed to replicate an entire pipeline, resulting in the coarse-grained baselines being unable to recover before the experiment ends. They will eventually recover as we see in~\figref{fig:autoscale} but only after suffering a period of 100\% SLO miss rate.

\subsection{Sensitivity Analysis}
\label{sec:eval:sensitivity}
We evaluate the sensitivity and robustness of the \planner and the \reactive.
We analyze the accuracy of the \estimator in estimating tail latencies from the sample trace and the \planner's response to varying arrival rates, latency SLOs, and burstiness factors.
We also analyze the \reactive's sensitivity to changes in the arrival process and ability to re-scale individual pipeline stages to maintain latency SLOs during these unexpected changes to the workload.

\begin{figure}[t]
	\centering
	\includegraphics[width=\columnwidth]{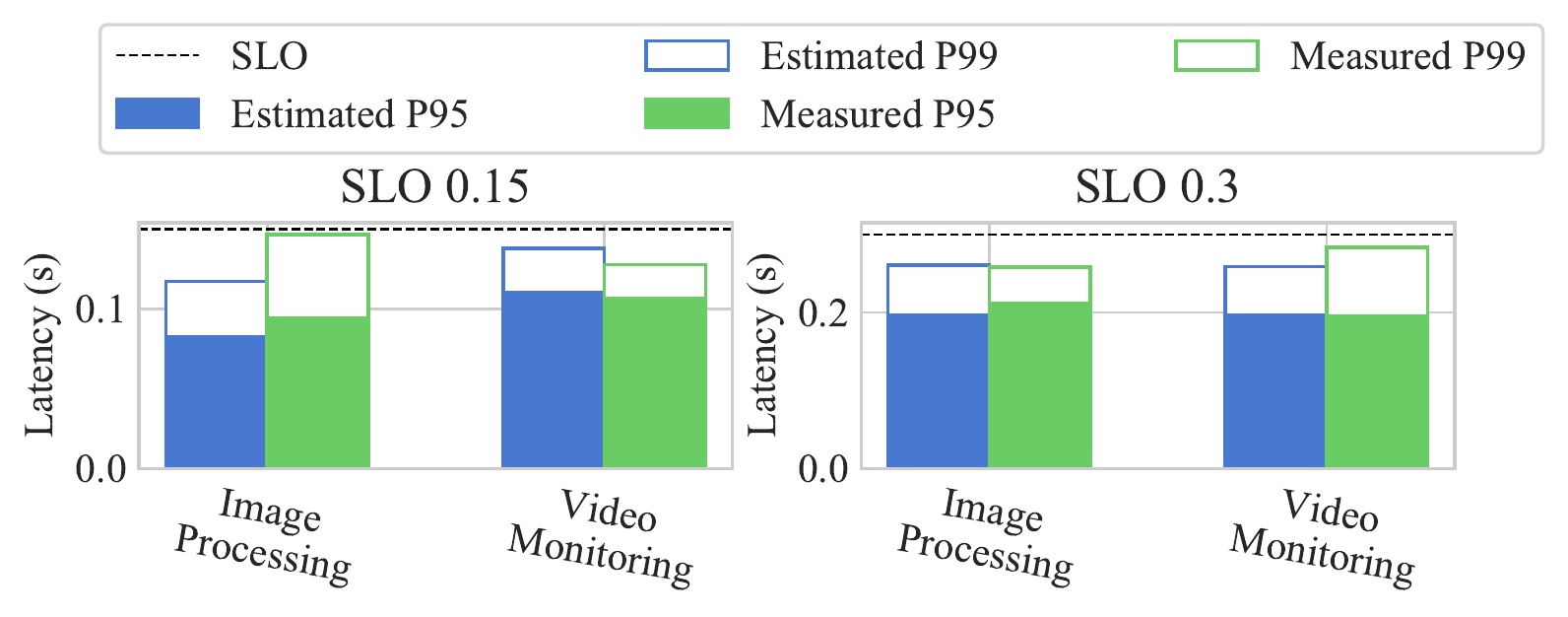}
	\vspace{-0.3in}
	\caption{
		\footnotesize{\textbf{Comparison of estimated and measured tail latencies.} We compare the latency distributions produced by the Estimator on a workload with $\lambda$ of 150 qps and CV of 4, observing that in all cases the estimated and measured latencies are both close to each other and below the latency SLO.
		}
	}
	\label{fig:estimatorfidelity}
	\vspace{-0.1in}
\end{figure}

\paragraph{Planner Sensitivity}
\label{sec:eval:sensitivity:planner}
We first evaluate how closely the latency distribution produced by the \estimator reflects the latency distribution of the running system in~\figref{fig:estimatorfidelity}.
We observe that the estimated and measured P99 latencies are close across all four experiments.
Further, we see that the Estimator has the critical property of ensuring that the P99 latency of feasible configurations is below the latency objective.
The near-zero SLO miss rates in~\figref{fig:exp:proactive} are a further demonstration of the Estimator's ability to detect infeasible configurations.

\begin{figure}[tb]
	\centering
	\includegraphics[width=\columnwidth]{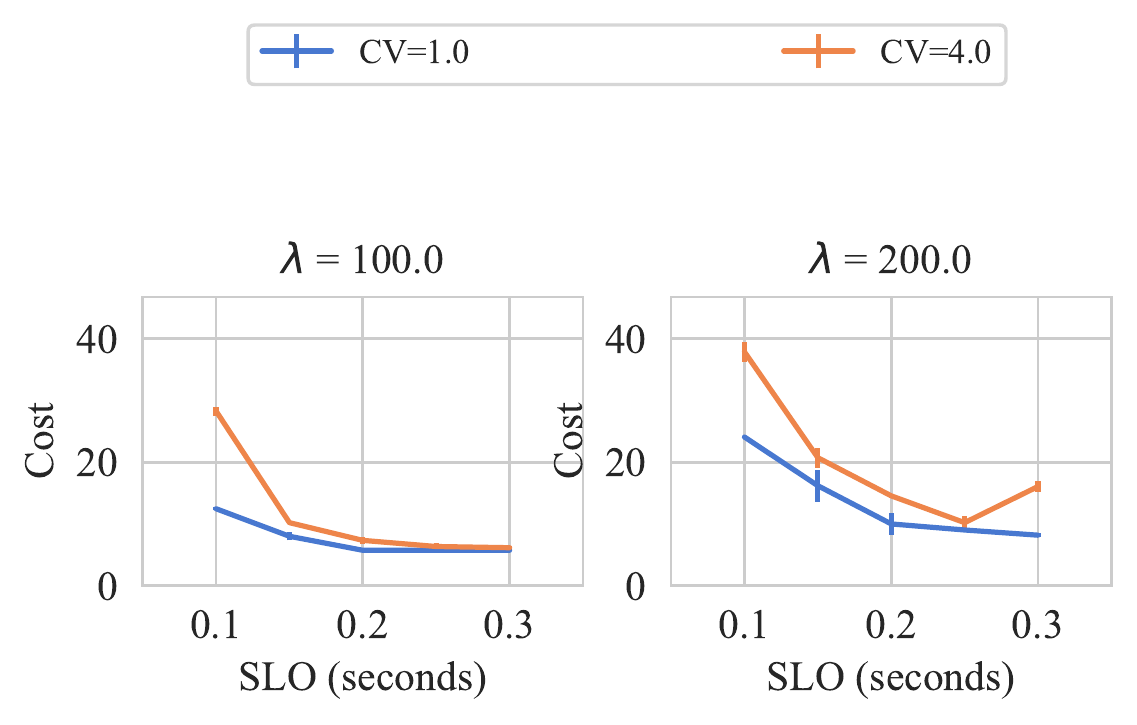}
	\vspace{-0.2in}
	\caption{
		\footnotesize{
			\textbf{\planner sensitivity:} Variation in configuration cost across different arrival processes and latency SLOs for the Social Media pipeline. We observe that 1) cost decreases as SLO increases, 2) burstier workloads require higher cost configurations, and 3) cost increases as $\lambda$ increases.}
	}
	\label{fig:opt-sensitivity}
	\vspace{-0.1in}
\end{figure}

Next, we evaluate the \planner's performance under varying load, burstiness, and end-to-end latency SLOs.
We observe three important trends in \figref{fig:opt-sensitivity}.
First, increasing burstiness (from CV=1 to CV=4) requires more costly configurations
as the Planner provisions more capacity to ensure that transient bursts
do not cause the queues to diverge more than the SLO allows.
We also see the cost gap narrowing between CV=1 and CV=4 as the SLO increases.
As the SLO increases, additional slack in the deadline can absorb more variability
in the arrival process and therefore fewer pipeline replicas are needed to process transient bursts within the SLO.
Second, the cost decreases as a function of the latency SLO.
While this downward cost trend generally holds,
the optimizer occasionally finds sub-optimal configurations, as it makes
locally optimal decisions to change a resource assignment. 
Third, the cost increases as a function of expected arrival rate, as more queries require more model replicas.

\paragraph{Tuner Sensitivity}
\label{sec:eval:sensitivity:arrival}
\label{sec:eval:sensitivity:burst}

\begin{figure}[t]
	\centering
	\includegraphics[width=\columnwidth]{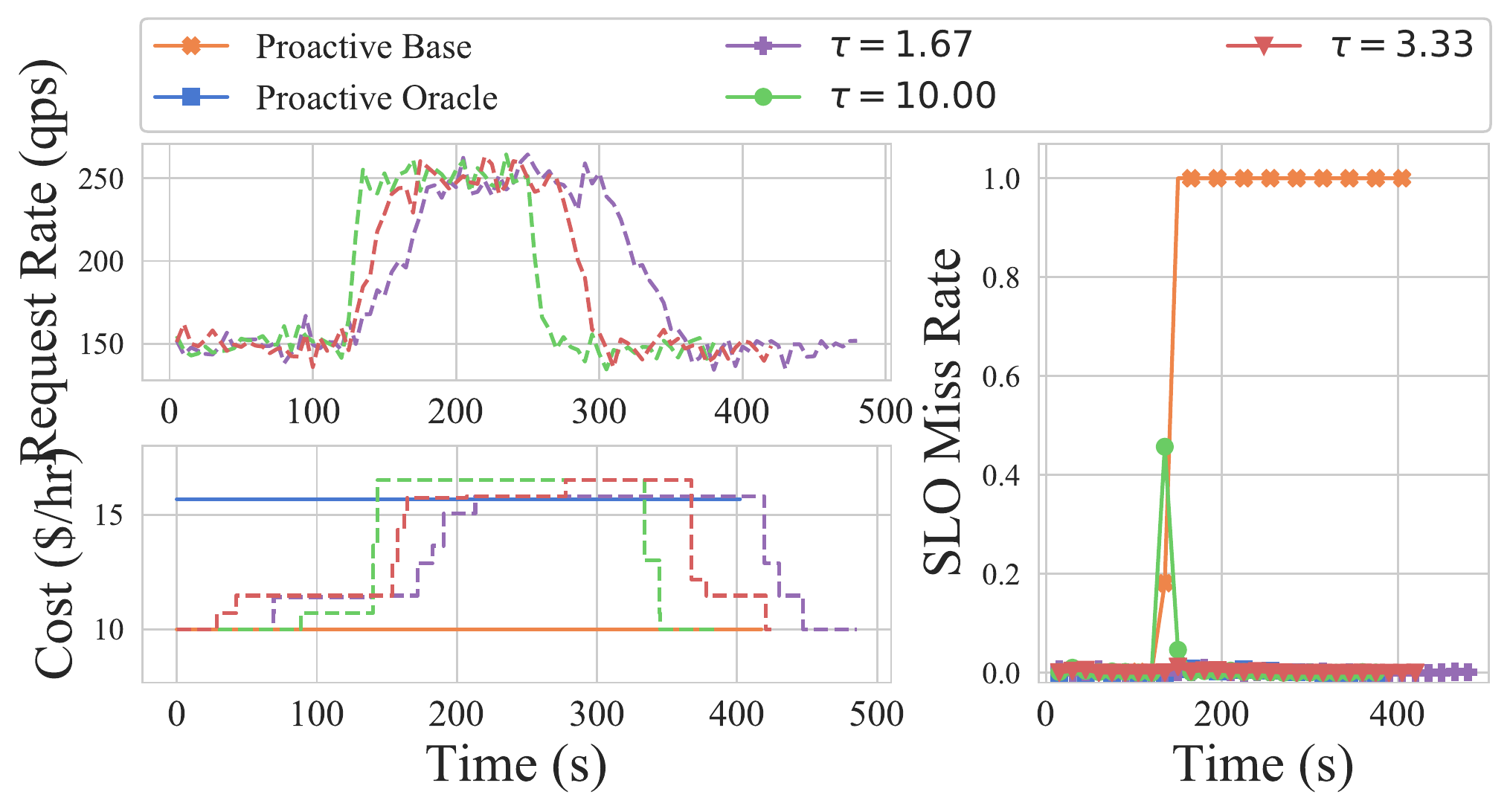}
	\vspace{-0.2in}
	\caption{
		\footnotesize{\textbf{Sensitivity to arrival rate changes (Social Media pipeline)}. We observe that the Tuner quickly detects and scales up the pipeline in response to increases in $\lambda$. Further, the Tuner finds cost-efficient configurations that either match or are close to those found by the Planner given full oracle knowledge of the trace.
		}
	}
	\vspace{-0.1in}
	\label{fig:eval:lambdasensitivity}
\end{figure}

A common type of unpredictable behavior is a change in the arrival rate.
We compare the behavior of \system with and without its Tuner enabled as the
arrival rate changes from the planned-for 150 QPS to 250 QPS.
We vary the rate of arrival throughput change $\tau$.
\system is able to maintain the SLO miss rate close to zero while matching or beating two alternatives: (a) a pipeline with only the Planner enabled but given full oracle knowledge of the arrival trace, and (b) a pipeline with only the Planner enabled and provided only the sample planning trace.
Neither of these baselines responds to changes in workload during live serving.
As we see in \figref{fig:eval:lambdasensitivity}, \system continues to meet the SLO, and increases the cost
of the pipeline only for the duration of the unexpected burst.
The oracle Planner with full knowledge of the workload is able to find the cheapest configuration at the peak because it is equipped with the ability to configure batch size and hardware type along with replication factor.
But it pays this cost for the entire duration of the workload.
The Planner without oracular knowledge starts missing latency SLOs as soon as the ingest rate increases as it is unable to respond to unexpected changes in the workload without the Tuner.

\begin{figure}[t]
\centering
    \centering
    \includegraphics[width=\columnwidth]{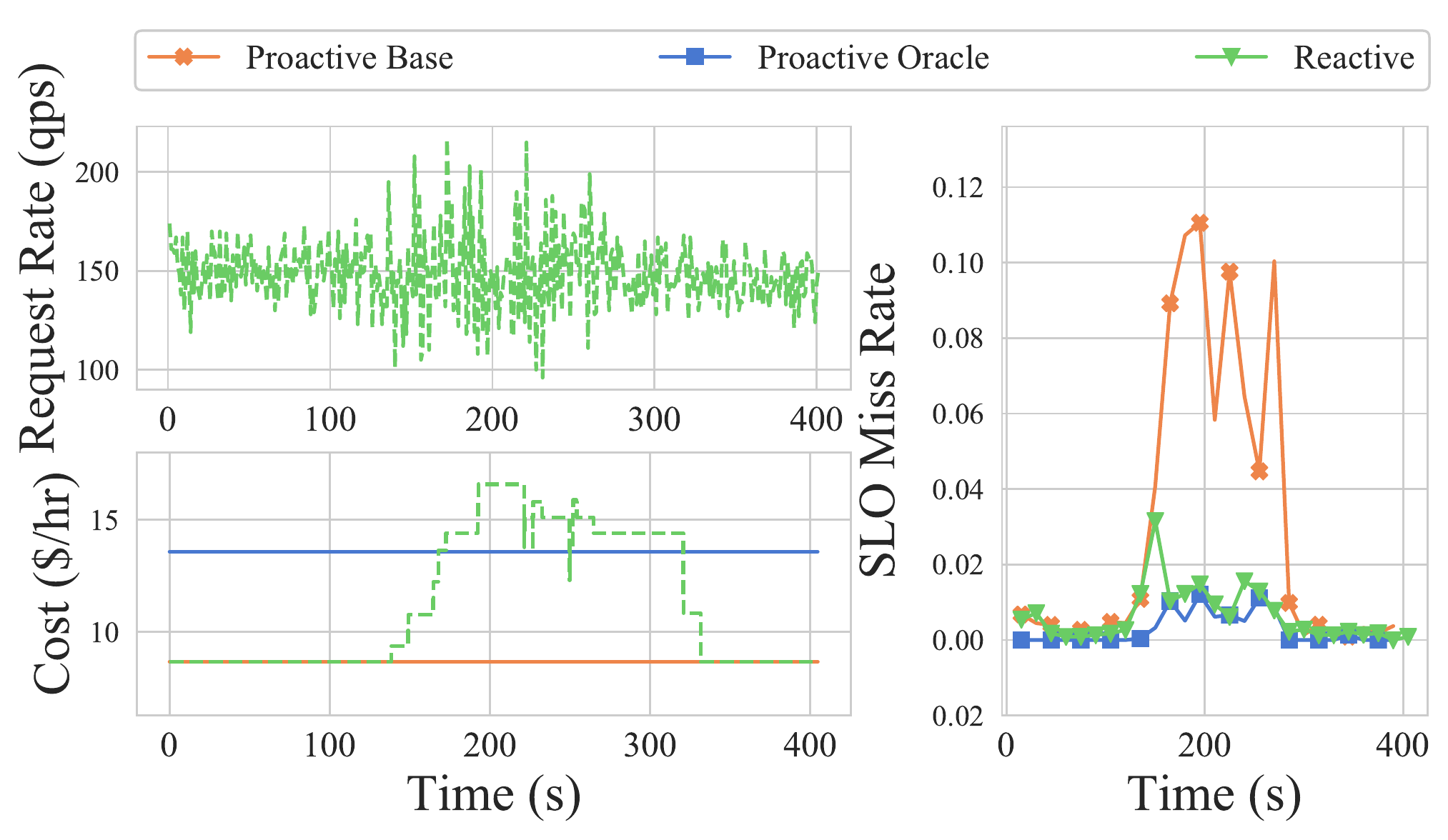}
\caption[short]{\footnotesize\textbf{Sensitivity to arrival burstiness changes (Social Media Pipeline).} We observe that the network-calculus based detection mechanism of the Tuner detects changes in workload burstiness and takes the appropriate scaling action to maintain a near-zero SLO miss rate.
}
\vspace{-0.1in}
\label{fig:eval:cvsensitivity}
\end{figure}

A less obvious but potentially debilitating change in the arrival process
is an increase in its burstiness, even while maintaining the same mean arrival rate $\lambda$.
This type of arrival process change is also harder to detect, as the common practice is to look at moments of the arrival rate distribution, such as the mean or 99th percentile. In~\figref{fig:eval:cvsensitivity} we show that \reactive is able to \emph{detect} deviation from expected arrival burstiness 
and react to meet the latency SLOs by employing the traffic-envelope detection mechanism described in~\secref{sec:reactive}.

\subsection{Attribution of Benefit}
\label{sec:eval:ablation}

\begin{figure}[t]
	\centering
	\includegraphics[width=\columnwidth]{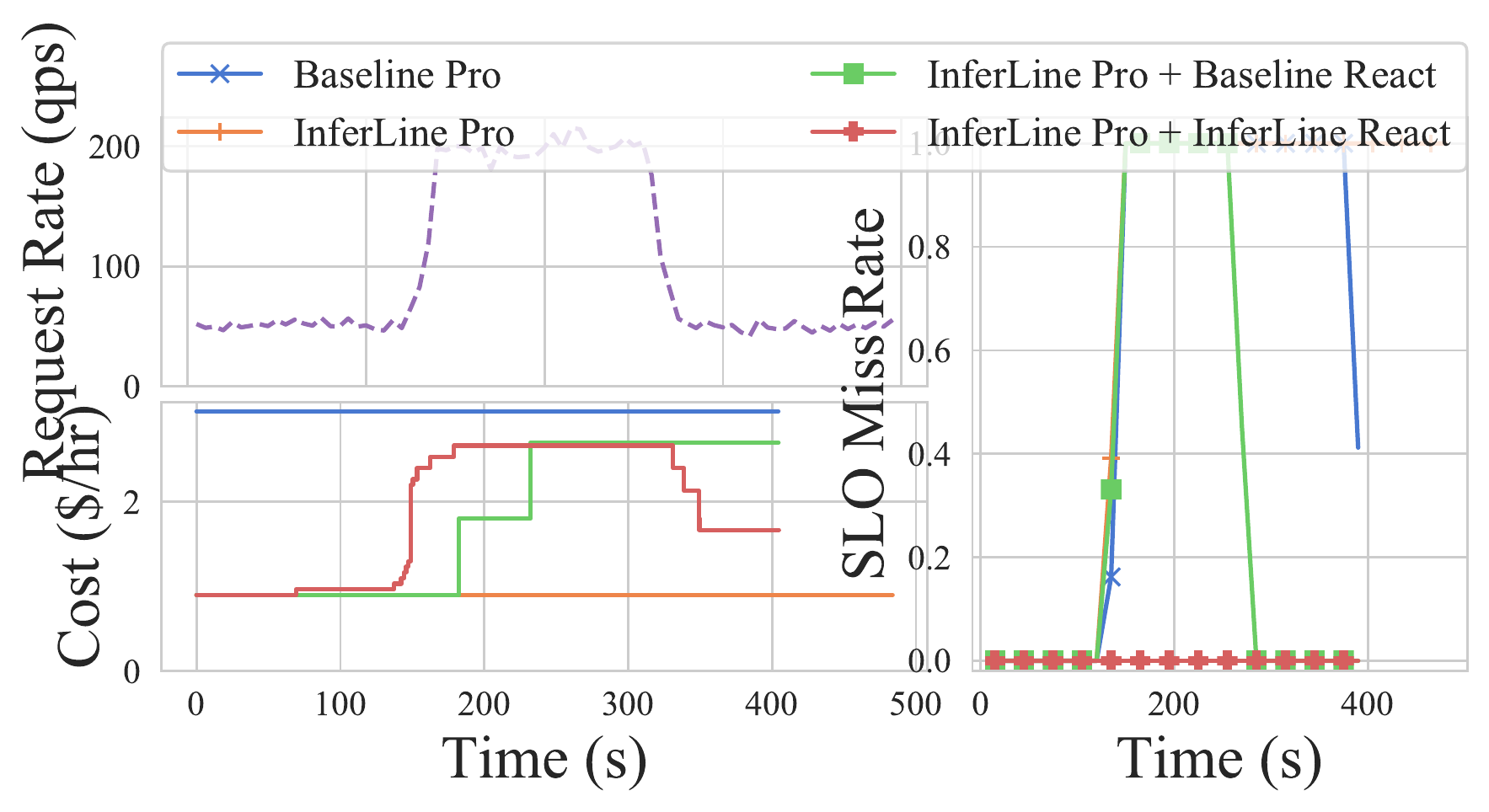}
	\vspace{-0.3in}
	\caption{
		\footnotesize{\textbf{Attribution of benefit between the \system low-frequency Planner and high-frequency Tuner on the Image Processing pipeline.} We observe that the Planner finds a more than 3x cheaper configuration than the baseline. We also observe that \system's Tuner is the only alternative that maintains the latency SLO throughout the workload.
		}
	}
	\label{fig:ablation}
	\vspace{-0.1in}
\end{figure}

\system benefits from 
(a) low-frequency planning and 
(b) high-freqency tuning. 
Thus, we evaluate the following comparison points:
baseline coarse grain planning (Baseline Plan), 
\system's planning (\system Plan),   
\system planning with baseline tuning (\system Plan + Baseline Tune), and 
\system planning with \system tuning (\system Plan + \system Tune), 
building up from pipeline-level configuration to the full feature set \system provides.
\system's Planner reduces the cost of the initial pipeline configuration by more than 3x~(\figref{fig:ablation}), but starts missing latency SLOs when the request rate increases.
Adding the baseline tuning mechanism (\system Plan + Baseline Tune) adapts the configuration, but too late to completely avoid
SLO misses, although it recovers faster than planning-only alternatives.
The \system Tuner has the highest SLO attainment and is the only alternative that maintains the SLO across the entirety of the workload.
This emphasizes the need for both the Planner for initial cost-efficient pipeline configuration, and the Tuner to promptly and cost-efficiently adapt to unexpected workload changes.


\subsection{Multiple Prediction-Serving Frameworks}
\label{sec:eval:gen}
\begin{figure}[t]
\centering
\includegraphics[width=\columnwidth]{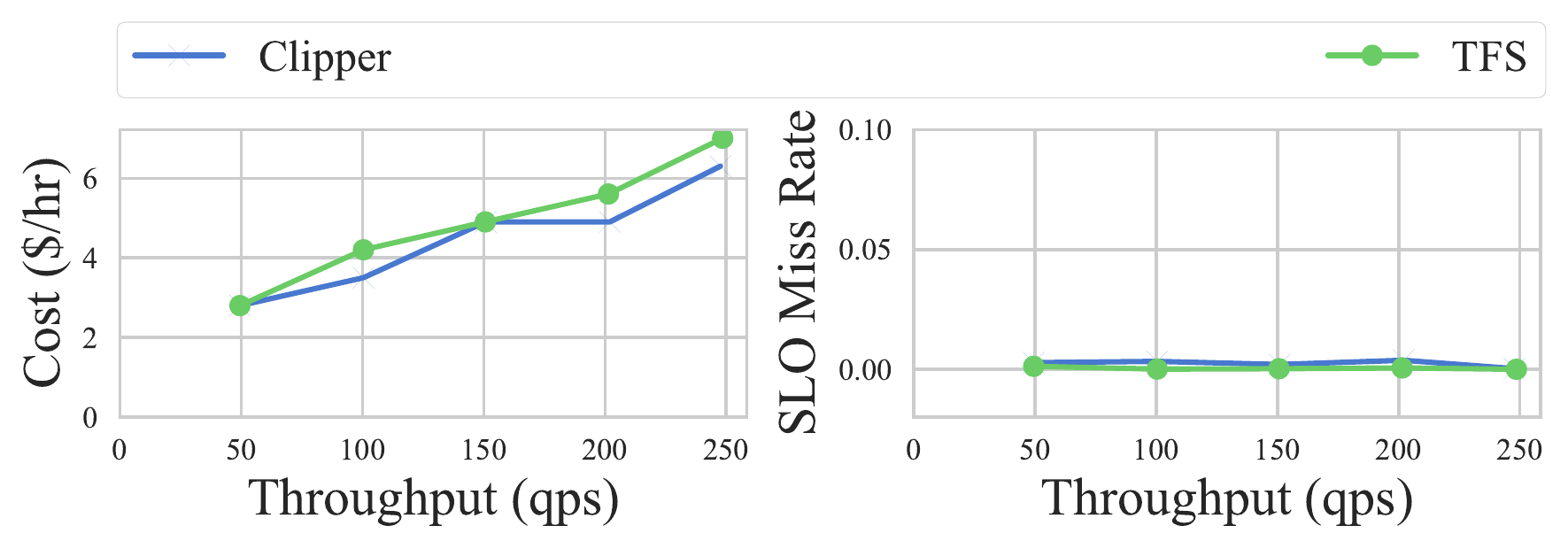}

\vspace{-0.1in}
\caption[short]{\footnotesize \textbf{Comparison of the \system Planner provisioning the TF Cascade pipeline in the Clipper and TensorFlow Serving (TFS) prediction-serving frameworks.} The SLO is 0.15 and the CV is 1.0.
}
\vspace{-0.2in}
\label{fig:eval:gen}
\end{figure}

The contributions of this work generalize to different underlying serving frameworks.
Here, we evaluate the \system \planner running on top of both Clipper and TensorFlow Serving (TFS).
In this experiment, we achieve the same low latency SLO miss rate for both prediction-serving frameworks.
This indicates the generality of the planning algorithms used to configure individual models in \system.
In \figref{fig:eval:gen} we show both the SLO attainment rates and the cost of pipeline provisioning when running \system on the two serving frameworks.
The cost for running on TFS is slightly higher due to some additional RPC serialization overheads not present in Clipper.

\section{Related Work}
\label{sec:prevwork}

A number of recent efforts study the design of generic prediction serving systems~\cite{clippernsdi17,Baylor17,tfserving,tensorrtserver}.
TensorFlow Serving~\cite{tfserving} is a commercial grade prediction serving system primarily designed to support 
prediction pipelines implemented using TensorFlow~\cite{tensorflow}, but does not provide any automatic provisioning or support latency constraints.
Clipper adopts a containerized design allowing each model to be individually managed, configured, and deployed in separate containers, but does not support prediction pipelines or reasoning about latency deadlines across models.
TensorRT Inference Server~\cite{tensorrtserver} from NVIDIA adopts a similar design to TensorFlow Serving and is optimized for NVIDIA GPUs.



Several systems have explored offline pipeline configuration for data pipelines~\cite{Jamshidi,Bilal}. However, these target generic data streaming pipelines.
They use black box optimization techniques that require running the pipeline end-to-end to measure the performance of each candidate configuration.
\system instead leverages performance profiles of each stage and a simulation-based performance estimator to explore the configuration space without needing to run the pipeline.

\begin{figure}[t]
	\begin{center}
		\subfigure[Increasing CV]{\includegraphics[width=0.3\columnwidth]{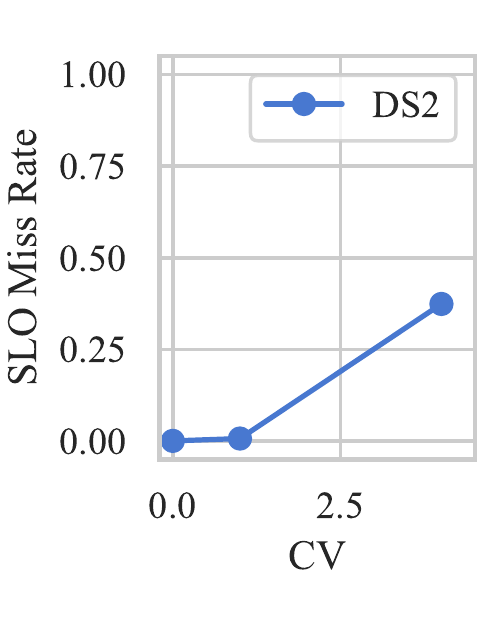}\label{fig:exp:ds2changingcv}
		}
		\subfigure[Increasing Arrival Rate]{\includegraphics[width=0.66\columnwidth]{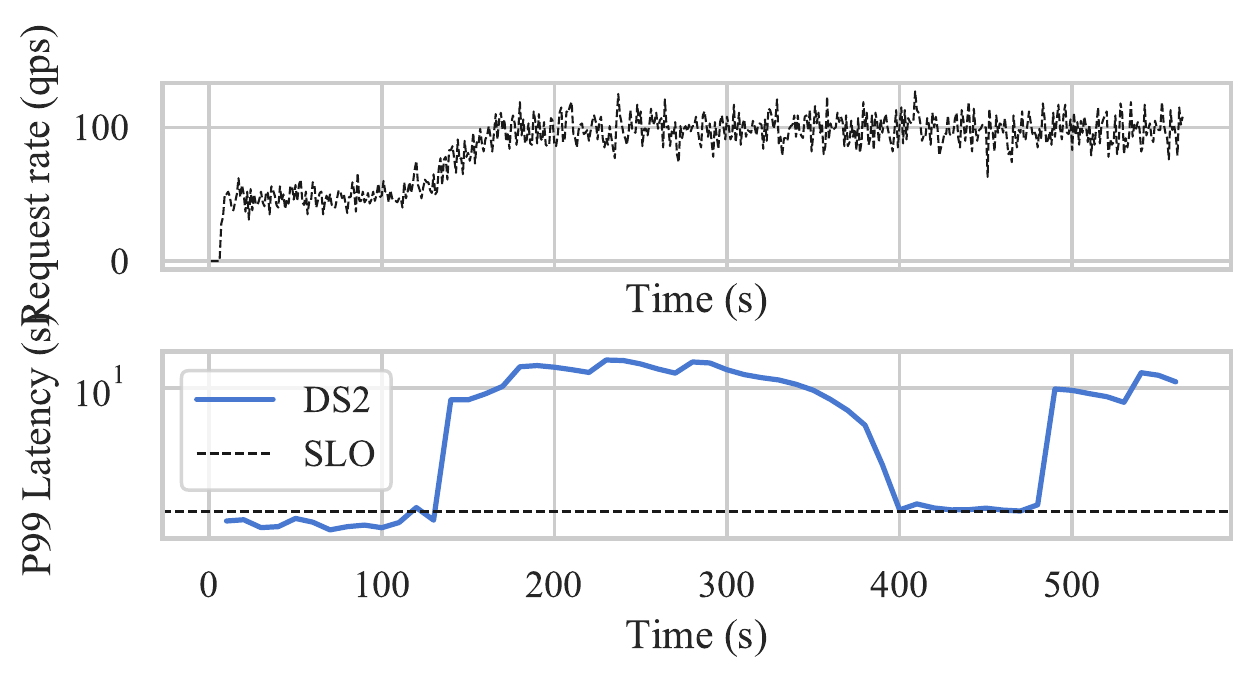}\label{fig:exp:ds2changinglambda}
		}
	\end{center}
	\vspace{-0.15in}
	\caption{
		\footnotesize \textbf{Performance of DS2 on Bursty and Stochastic Workloads.} We observe that DS2 is unable to maintain the latency SLO under (a) bursty workloads, and (b) workloads with increasing request rate.
	}
	\label{fig:exp:ds2}
	\vspace{-0.15in}
\end{figure}

Dynamic pipeline scaling is a critical feature in data streaming systems to avoid backpressure and overprovisioning.
Systems such as~\cite{Kalavri,Floratou} are throughput-oriented with the goal of maintaining a well-provisioned system under changes in the request rate.
The DS2 autoscaler in ~\cite{Kalavri} estimates true processing rates for each operator in the pipeline by instrumenting the underlying streaming system.
They use these processing rates in conjunction with the pipeline topology structure to estimate the optimal degree of parallelism for all operators at once.
In contrast, ~\cite{Floratou} identifies a single bottleneck stage at a time, taking several steps to converge from an under-provisioned to a well-provisioned system.
Both systems provision for the average ingest rate and ignore any burstiness in the workload which can transiently overload the system.
In contrast, \system maintains a traffic envelope of the request workload and uses this to ensure that the pipeline is well-provisioned for the peak workload across several timescales simultaneously, including any burstiness (see ~\secref{sec:reactive}).

In~\figref{fig:exp:ds2} we evaluate the performance of DS2~\cite{Kalavri} on its ability to meet latency SLOs under a variety of workloads.
We deployed the Image Processing pipeline (\figref{fig:exp:heavycpu}) in DS2 running on Apache Flink~\cite{flink} without any batching.
As we can see in~\figref{fig:exp:ds2changingcv}, provisioning for the average request rate is sufficient to meet latency objectives under uniform workloads.
But as CV increases to 4.0, the latency SLO miss rate increases
as bursts in the request rate transiently overload the system, causing queueing delays until the system recovers.
In addition, DS2 occasionally misinterprets transient bursts as changes in the overall request rate and scales up the pipeline, requiring Apache Flink to halt processing and save state before migrating to the new configuration.

We observe this same degradation under non-stationary workloads in~\figref{fig:exp:ds2changinglambda} where we measure P99 latency over time for a workload that starts out with a CV of 1.0 and a request rate of 50 qps, then increases the request rate to 100 qps over 60 seconds.
It takes nearly 300 seconds after the request rate increase for
the system to re-stabilize and the queues to fully drain from the repeated pipeline re-configurations.
In contrast, as we see in~\figref{fig:eval:lambdasensitivity} and~\figref{fig:eval:cvsensitivity}, \system is capable of maintaining SLOs under a variety of changes to the workload dynamics.

A few streaming autoscaling systems consider latency-oriented performance goals~\cite{fu:drs2017,lohrmann15}.
The closest work to \system, ~\cite{lohrmann15} from Lohrmann et al. as part of their work on Nephele ~\cite{nephele},
treats each stage in a pipeline as a single-server queueing system and uses queueing theory to estimate the total queue waiting time of a job under different degrees of parallelism.
They leverage this queueing model to greedily increase the parallelism of the stage with the highest queue waiting time until they can meet the latency SLO.
However, their queueing model only considers average latency, and provides no guarantees about the behavior of tail latencies.
\system's Tuner automatically provisions for worst-case latencies.


VideoStorm~\cite{VideoStorm} explores the design of a streaming video processing system
that adopts a distributed design with pipeline operators provisioned across compute nodes
and explores the combinatorial search space of hardware and model configurations. 
VideoStorm jointly optimizes for quality and lag and does not provide latency guarantees.

Nexus~\cite{nexussosp19} is a recent system configures DNN inference pipelines for video-streaming applications.
Similar to \system, it uses model profiles to understand model batching behavior and provisions pipelines for end-to-end latency objectives.
However, they do not configure which hardware to run models on, instead assuming a homogenous cluster of GPUs.
Furthermore, they rely on admission control to reject queries during transient spikes while \system's Tuner quickly re-scales the pipeline to maintain SLOs without rejecting  queries.


A large body of prior work leverages profiling for scheduling, including recent work on workflow-aware scheduling~\cite{rodrigo-hpdc17,morpheus-osdi16}.
In contrast, \system{} exploits the compute-intensive and side-effect free nature of ML models to estimate end-to-end pipeline performance based on individual model profiles.

Autoscale~\cite{autoscale} comprehensively surveys work aimed at automatically scaling the number of servers 
reactively, subject to changing load in the context of web services. Autoscale works well for single model replication without batching as it assumes bit-at-a-time instead of batch-at-a-time query processing.
However, we find that the \system high-frequency Tuner outperforms the coarse-grain baselines using the Autoscale scaling mechanism on both latency SLO attainment and cost (\secref{sec:eval:e2e:reactive}).





\section{Limitations and Generality}
\label{sec:limitations}

One limitation of the Planner is its assumption that the available hardware has a total ordering of latency across all batch sizes.
As specialized accelerators for ML continue to proliferate, there may be settings where one accelerator is slower than another at smaller batch sizes but faster at larger batch sizes.
This would require modifications to the hardware downgrade portion of the planning algorithm to account for this batch-size dependent ordering.

A second limitation is the assumption that the inference latency of ML models is independent of their input.
There are emerging classes of machine learning tasks where state-of-the-art models have inference latency that varies based on the input.
For example, object detection models ~\cite{faster-rcnn, yolo} will take longer to make predictions on images with many objects in them.
One way of modifying \system to account for this is to measure this latency distribution during profiling based on the variability in the sample queries and use the tail of the distribution (\eg 99\% or $k$ standard deviations above the mean) as the processing time in the estimator, which will lead to feasible but more costly configurations.

Finally, while we only study machine learning prediction pipelines in this work, there may be other classes of applications that have similarly predictable performance and can therefore be profiled.
We leave the extension of \system to applications beyond machine learning as future work.

\section{Conclusion}
\label{sec:conclusion}
In this paper we studied the problem of provisioning and managing prediction pipelines to meet end-to-end tail latency requirements at low cost and across heterogeneous parallel hardware.
We introduced \system--a system which efficiently provisions prediction pipelines subject to end-to-end latency constraints.
\system combines a low-frequency Planner that finds cost-optimal configurations with a high-frequency Tuner that rapidly re-scales pipelines to meet latency SLOs in response to changes in the query workload.
The low-frequency Planner combines profiling, discrete event simulation, and constrained combinatorial optimization to find the cost minimizing configuration that meets the end-to-end tail latency requirements without ever instatiating the system (\secref{sec:offline}).
The high-frequence Tuner uses network-calculus
to quickly auto-scale each stage of the pipeline to accommodate changes in the query workload (\secref{sec:online}).
In combination, these components achieve the combined effect of \emph{cost-efficient} heterogeneous prediction pipeline provisioning that can be deployed to a variety of prediction-serving frameworks to serve
applications with a range of tight end-to-end latency objectives.
As a result, we achieve up to 7.6x improvement in cost and 34.5x improvement in SLO attainment for the same throughput and latency objectives over 
state-of-the-art provisioning alternatives.

\newpage

\bibliographystyle{abbrv}

{\small
\bibliographystyle{acm}
\bibliography{references}
}
\end{document}